\documentstyle[aps,prl,eqsecnum,preprint,tighten]{revtex}
\begin{document}
\title{Edge excitations of paired fractional quantum Hall states}
\author{M. Milovanovi\'{c} and N. Read}
\address{Departments of Physics and Applied Physics, P.O. Box 208284\\
Yale University, New Haven, CT 06520}
\date{February 20, 1996}
\maketitle
\begin{abstract}
The Hilbert spaces of the edge excitations of several ``paired'' fractional 
quantum Hall states, namely the Pfaffian, Haldane-Rezayi and 331 states, 
are constructed and the states at each angular momentum level are enumerated. 
The method is based on finding {\em all} the zero energy states for those 
Hamiltonians for which each of these known ground states is the exact, 
unique, zero-energy eigenstate of lowest angular momentum in the disk 
geometry. For each state, we find that, in addition to the usual bosonic 
charge-fluctuation excitations, there are fermionic edge excitations. The 
wavefunctions for each case have a similar form, related to Slater 
determinants, and the edge states satisfy a ``projection rule'', that the 
parity of the number of fermions added to the edge equals the parity of the 
charge added. The edge states can be built out of quantum fields that 
describe the fermions, in addition to the usual scalar bosons (or Luttinger
liquids) that describe the charge fluctuations. The fermionic fields in the 
Pfaffian and 331 cases are a non-interacting Majorana (i.e.,\ real Dirac) 
and Dirac field, respectively. For the Haldane-Rezayi state, the field is 
an anticommuting scalar. For this system we exhibit a chiral Lagrangian that 
has manifest SU(2) symmetry but breaks Lorentz invariance because of the 
breakdown of the spin statistics connection implied by the scalar nature of 
the field and the positive definite norm on the Hilbert space. Finally we 
consider systems on a cylinder where the fluid has two edges and construct 
the sectors of zero energy states, discuss the projection rules for combining 
states at the two edges, and calculate the partition function for each edge 
excitation system at finite temperature in the thermodynamic limit. The 
corresponding theory for the hierarchy and its generalizations is also given. 
It is pointed out that the conformal field theories for the edge states are 
examples of orbifold constructions. Two appendices contain technical details.
\end{abstract}
\pacs{PACS: }


\section{Introduction}
\label{introduction}

The theory of the excitations at the edge of an incompressible fractional
quantum Hall state \cite{book} has undergone extensive development since its 
beginnings a few years ago \cite{wen,stone,haldane,wen2}. In the integer
quantum Hall effect, that is, when the bulk fluid fills an integer number of
Landau levels, the edge excitations are essentially single electrons occupying
single-particle edge states \cite{halp82}, that propagate in one direction
along the edge and correspond to the classical skipping orbits. There is one 
``channel'' of such edge states for each filled Landau level; each channel can 
be considered as a more or less non-interacting, one-dimensional,
uni-directional Fermi sea. In the fractional effect, the edge
excitations, like the bulk states, are highly correlated and cannot be
described by single-electron states. The basic variables are density
fluctuations which propagate in one direction along the edge. The quantum field
theory which describes these is a chiral Luttinger liquid \cite{wen,wen2}. 
In the simplest case, that of the Laughlin states at filling factors $1/q$, $q$
odd, this density mode is the only low-energy excitation at the edge. In the 
special case of the integer effect at filling factor 1, this is equivalent, via
bosonization, to the single-electron, Fermi-sea description \cite{stone}. The
theory was soon extended \cite{wen,wen2} to the edge excitations of the 
hierarchy theory \cite{book}, which yields an incompressible ground state in 
the bulk for all rational filling factors $p/q$ with $q$ odd. In general, some 
chiral conformal field theory, that generalizes the chiral Luttinger liquid, is 
expected to describe the low-energy, long-wavelength excitations.
It was predicted in Ref.\ \cite{mr} that this theory would in general 
be the same as the conformal field theory whose correlation functions 
reproduce the bulk wavefunctions, and which describes such universal properties
of the bulk states as the statistics of their fractionally-charged excitations,
as it does for the hierarchy states. 
This deep connection implies that the properties of the edge states are not
only of interest in their own right, but they can also be used to probe the
properties of the underlying bulk state. Effects in tunneling into or between
edge states have been the subject of various works \cite{wentun,kanefish}.

In this paper we wish to extend the theory of edge states to cover
some other interesting states that have been proposed and which do not fit 
into the hierarchy scheme. In particular, there are (i) the Haldane-Rezayi (HR)
\cite{hr} state,
proposed to explain the plateau observed at $\nu=5/2$ in terms of a half 
filling of the first excited Landau level, in which the electrons have no net
polarization; (ii) the Moore-Read (Pfaffian) state \cite{mr}, again 
for a half-filled Landau level but this time with spin-polarized or spinless 
electrons. They are ground states of
electron systems with special short range interactions, described later in this
paper. The nature of the
observed 5/2 state remains controversial (other suggestions include an 
alternative spin-singlet state \cite{belkhir}, which we believe to be a
spin-singlet generalized hierarchy state), while the Pfaffian has been 
proposed as an explanation for a $\nu=1/2$ plateau in double layer systems 
\cite{gww}, though theoretical calculations do not support this suggestion. 
Instead, they suggest \cite{he} that the ground state there is (iii) 
the so-called 331 state. Since this is constructed as a two-component
generalization of the Laughlin states, it is part of the generalized hierarchy 
\cite{read90},
but since it can also be interpreted, like the HR and Pfaffian states,  
as a ``paired'' state, it will be natural to include it here. 

The terminology ``paired'' state must be interpreted carefully. It is an old
idea that Laughlin's states might be generalized if the electrons are first
grouped into clusters of $m$ particles (such as pairs, $m=2$) and the 
resulting objects of charge $m$ then form a Laughlin state. While this may be
possible, and was apparently part of the idea of HR \cite{hr}, it is not quite
what we have in mind. In terms of the now-popular composite fermions, which
consist of an electron plus an even number $q$ of attached vortices, and 
which at filling factor $1/q$ behave as particles in zero net field \cite{hlr},
the paired states are obtained by forming a BCS-type paired ground-state 
wavefunction, rather than a Fermi sea (the real-space wavefunctions are given 
in \S II). This was first pointed out in Ref.\ \cite{mr}, and has been 
discussed more recently in Refs.\ \cite{halpnewport,ho}. We note that it is 
not clear 
that these two procedures lead to equivalent states. In particular, the first 
idea seems to lead to a prediction of abelian statistics for fractionally 
charged quasiparticles \cite{gww}, while the second has been connected with 
nonabelian statistics \cite{mr}. The comparison of the two procedures hinges 
on the question whether the two operations, of grouping particles into pairs, 
and of attaching an appropriate number of vortices, commute. In any case, they 
do both lead to the result that quasiparticles have charges in multiples of 
$1/2q$ rather than $1/q$ \cite{mr}, and, in the latter procedure, to the 
existence of BCS-type (composite) fermionic excitations obtained by breaking 
pairs; these are expected to have a gap in their spectrum. In this paper, we 
will see that the gap for the latter excitations goes to zero 
at the edge, and the fermions appear as gapless edge excitations, in addition 
to the usual bosonic charge fluctuations. The fermions can be described by 
quantum field theories which are related to relativistic conformal field 
theories (CFTs); in the cases of the Pfaffian and 331 states, these are the 
chiral versions of familiar Majorana and Dirac fermions, respectively. 

The goal of this paper is to understand the structure of the Hilbert spaces,
and the field theories, of the edge excitations of these paired states. 
Only closed systems are considered, with the fluid in the form of either a 
droplet with one edge, or an annulus with two oppositely-moving edges.
Knowledge of these field theories provides the necessary background for study
of the tunneling and other properties of these states, which might be useful as
a diagnostic for the nature of the bulk ground state. 

There is some previous work on the edge states considered here. Wen 
\cite{wen3} has provided numerical evidence for decoupled Majorana fermions 
at the edge of the Pfaffian state. Wen, Wu and Hatsugai \cite{wwh} studied 
the edge excitations of the HR and other states, using techniques developed by 
Wen and Wu \cite{ww} for applying operator product expansions in CFT to FQHE
wavefunctions along the lines of \cite{mr}. They obtained analytical results
for the edge states, but were not able to show either the completeness or the
linear independence of their states, though the dimensions of the spaces were
confirmed numerically for the lowest excitations. 

In the remainder of this paper we will first write down (in \S II), for the 
Pfaffian, HR and 331 states, wavefunctions for edge excitations
that are zero energy eigenstates of the appropriate Hamiltonian and manifestly 
allow an interpretation as decoupled systems of fermion and the usual boson
excited states at the edge. These states appear to be linearly independent, and
it remains to check that we have obtained all the edge (or zero energy) states.
This is proved in Appendix A. Linear independence is confirmed, up to the 
eighth level of excited states for the Pfaffian, and the sixth level for HR,
by direct construction in Appendix B. For the Pfaffian, this reproduces and 
extends Wen's \cite{wen3} numerical results, and analogous results for the HR 
state by Wen, Wu and Hatsugai \cite{wwh}. Our method has the advantage over 
the computer calculations of being {\em analytic} and valid for any number of 
electrons. In \S III, we present a $1+1$ fermion field theory for the neutral 
part of the spectrum of the HR state; the SU(2) symmetry of the system is 
explicit in this construction. The exponents in the singularity in the electron 
occupation number at the edge are predicted. We also present a CFT whose 
correlators reproduce the {\em bulk} wavefunctions in the manner 
of \cite{mr}. Finally, in \S IV, we consider systems on a cylinder, with two 
edges, which gives further information about the structure of the systems. This 
information, a complete description of the number of states at each excitation 
level in the thermodynamic limit, is conveniently expressed as a partition 
function similar to the usual grand canonical one. The results are interpreted 
using the CFT idea of an orbifold construction. We emphasize that our basic
results are derived without use of CFT, which is needed only in \S III and at 
the end of \S IV, where the results are compared with CFTs. 

\section{Edge States of a disk}
\subsection{Edge states of the Laughlin state}

In this section we will first review what is known about the
edge states of the Laughlin states, emphasizing points  
that will help us in studying other FQHE systems. We then turn to results
for the Pfaffian, HR and 331 states.

Considering first the interior of a system, i.e.\ in the thermodynamic
limit where the edge disappears to infinity, or in a system filling a finite 
but closed geometry such as the sphere, a FQH system at a given rational 
filling fraction possesses by definition a unique ground state (except for 
some global phenomena in the case of surfaces of nontrivial topology) and a 
gap for all excited states of the same or  
higher density than this ground state. In a number of interesting cases 
(with the particles confined to a fixed Landau level, usually the lowest) 
a model short-range repulsive interaction can be found for which this 
ground state has a wavefunction that is known exactly, and is a zero 
energy eigenstate. Therefore, turning to a QH droplet in a plane, 
with these interactions there is always a possibility of excitations of zero 
energy which are ``inflations'' of the densest zero energy state of the 
system. Hence a zero energy state always involves some deformation of the edge
compared with the ground state. 
 
The simplest examples of these systems are modeled in terms of Haldane's 
pseudopotentials \cite{hald} by the Hamiltonian 
(we work in the lowest Landau level throughout)  
\begin{equation}
   H= \sum_{l=0}^{q-1}{\cal V}_{l}\sum_{i<j}{\cal P}_{l}^{ij}
\label{pseudH}
\end{equation}
where ${\cal V}_{l}$ are positive constants (pseudopotentials) and 
${\cal P}_{l}^
{ij}$ is the projection operator onto the relative angular momentum state of
angular momentum $l$ for particles $i$ and $j$.  The densest zero energy 
eigenstate of $H$ (\ref{pseudH}), that is the one with the lowest total
angular momentum (= total degree of the polynomial part) is the Laughlin state
(here in the symmetric gauge)
\cite{laugh} 
\begin{equation}
\Psi_{L}(z_{1},\ldots,z_{N})=\prod_{i<j}(z_{i}-z_{j})^{q}\exp[-\frac{1}{4}
\sum|z_i|^2].
\end{equation}
If the Laughlin state is multiplied by any symmetric polynomial in the 
$z_i$'s, it will still be a zero energy state \cite{hald}. We will elaborate a
little on this point.
                    
The Laughlin quasihole operator
\begin{equation}
   \prod_{i=1}^{N}(z_{i}-w),
\end{equation}
multiplied into $\Psi_L$, produces a quasihole located at $w$, and
many-quasihole states can be obtained by repeated use of this operator. All
such states are clearly zero energy excitations for (\ref{pseudH}), because the
wavefunction still vanishes as $(z_i-z_j)^q$ as $z_i\rightarrow z_j$.
The operator can be viewed as a generating function, through the expansion 
\begin{equation} 
   \prod_{i=1}^{N}(z_{i}-w)=e_{N}+\cdots+(-w)^{N-2}e_{2}+(-w)^{N-1}e_{1}
+(-w)^{N},
\label{qhexpn}
\end{equation}
for the elementary symmetric polynomials,
\begin{equation}
e_{n}=\sum_{1\leq i_1<i_2<\cdots<i_n\leq N}z_{i_{1}}\cdots z_{i_{n}}.
\end{equation}    
The symmetric polynomials in the
$z_i$ form a closed set under the operations of taking sums, differences and
products, so they form a ring with unit element 1; allowing linear combinations
with arbitrary complex coefficients, they form an algebra. 
All such polynomials, and therefore all zero energy states, are obtained 
as linear combinations of
 products of the $e_n$, {\it i.e.} the $e_n$'s form a basis for the algebra of
symmetric polynomials. 
The $e_n$'s can be obtained by integrating the quasihole operator, times a
suitable factor, over all $w$;
thus all zero energy states can also be obtained as
linear combinations of products of integrals 
of quasihole operators acting on $\Psi_L$.
Single Laughlin quasihole states reconstructed from the $e_n$'s {\it via} 
(\ref{qhexpn}) 
are not orthogonal and may be considered as 
coherent states, similar to those of a single electron in the lowest Landau
level, which can be expanded in angular momentum eigenstates. Likewise here,
the $e_n$ are operators that increase the angular momentum of the electrons by
$n$. Also, as operators, they clearly commute.

As long as $w$ lies well inside the disk of radius $\sqrt{2q(N-1)}$ formed by 
the electrons, say more than one magnetic length from the edge, it represents 
the position of the quasihole. As $w\rightarrow\infty$, the disturbance in
density due to the quasihole approaches the edge and eventually, in the limit,
becomes trivial. The leading corrections to this limit are then the terms
$e_1$, $e_2$, \ldots\ so that $e_n$ with larger angular momentum represent
larger distortions of the edge. This is borne out by the energetics, if we
introduce a term $\lambda M$ into the Hamiltonian, where $M$ is the operator
representing total angular momentum about the origin. Since rotations are a 
symmetry of the unperturbed problem, this term merely splits the degeneracy of 
the different angular momentum eigenstates. Then it is clear that states 
$e_n\Psi_L$ have energy increasing linearly with $n$ and since gapless 
excitations can occur only near the edge of an incompressible QH state, we can 
again conclude that these states are edge excitations. This can be verified in 
detail for the $q=1$ case which is a filled Landau level \cite{stone,wen}.
The same expansion for these coherent states with more than one quasihole
gives for each linearly independent bosonic state a corresponding edge
state. By the arguments above, these must span the full space of edge
excitations. 

One may count the number of states at each increased angular momentum, as
follows. Note that to describe {\em edge} states, we consider the limit
$N\rightarrow\infty$, with $\Delta M=M-M_0$ fixed ($M_0$ is 
the angular momentum in the ground state, and $M_0=qN(N-1)/2$ in the Laughlin
state). Bulk states are obtained either by fixing $w$'s in quasihole states 
or by applying $e_n$'s with $n$ of order $N$, and then $N\rightarrow \infty$ 
in either case. Since $\Delta M$ in an edge state $\prod_\alpha
e_{n_\alpha}\Psi_L$ is $\Delta M = \sum_\alpha n_\alpha$, the total number of
states at $\Delta M$ is $p(\Delta M)$, the number of ways $\Delta M$ can be 
divided (``partitioned'') into
positive integer parts whose sum is $\Delta M$. However, their 
meaning is clearest if we use a different basis for the algebra of symmetric 
polynomials, namely the sums of powers
\begin{equation}
s_n=\sum_i z_i^n
\label{sn}
\end{equation}
(these are not all independent for finite $N$, but must all be used as
$N\rightarrow\infty$), which are one body operators, and, up to
constant factors, can be viewed as the 
positive angular momentum components of the change in density at the edge from
the ground state, projected into the space of zero energy states. Thus the edge 
states are built up out of repeated applications of density operators which do
behave as boson creation operators (i.e.~they commute) for $n$ 
positive. 
The components of the projected density with $n$ negative act as destruction 
operators and with a suitable normalization the algebra of independent simple 
harmonic oscillators is obtained, or equivalently, the abelian (U(1)) analogue 
of a Kac-Moody algebra.
This point of view is extensively discussed in \cite{wen,stone,wen2}.

The partition function can be obtained from Euler's generating function which 
is an infinite series in an indeterminate $x$:
\begin{equation}
{\cal Z}(x)\equiv 1+\sum_{\Delta M=1}^\infty p(\Delta M)x^{\Delta M}\equiv
\prod_{n=1}^\infty(1-x^n)^{-1},
\end{equation}
where we used the binomial expansion
\begin{equation}
(1-x^n)^{-1}=1+x^n+x^{2n}+\ldots.
\end{equation}
We recognize ${\cal Z}(x)$ as the statistical-mechanical partition 
function of a
chiral Bose field, that is a collection of simple harmonic oscillators of
frequencies $n\omega$, $n=1$, $2$, \ldots, if we set $x=\exp (-\omega/k_B T)$ 
(and $\hbar=1$).
For the convenience of the reader at later points in this paper we include a
table of the partition function $p(\Delta M)$ for small $\Delta M$:

\begin{center}
{\begin{tabular}{c|c|c|c|c|c|c|c|c}
	$\Delta M $  & 1 & 2 & 3 & 4 & 5 & 6 & 7 & 8   \\ \hline
       dim=$p(\Delta M)$ & 1 & 2 & 3 & 5 & 7 & 11 & 15 & 22 
\end{tabular}}\quad.
\end{center}

In addition to these states, we can obtain states with a net charge added to
the edge, either by changing the electron number, which of course can only give
integral charges, or by adding quasiholes at the center of the disk, which
allows the charge effectively added to the edge to be a multiple of $1/q$. The
wavefunctions of such a state, with a positive charge $m/q$ added at
the edge, are the same as the above except that a factor $\prod_i z_i^m$ is
included. The partition function for the states in each such sector is the same
as for $m=0$; these again represent density fluctuations on top of the state
which now has charge $m/q$ added. The states with different electron number
will, in the following, generally be found to play a role in the structure of
the theory. Whether the states with quasiholes added at the center should be
viewed as part of the edge theory is somewhat a matter of taste; they could 
alternatively be viewed as a part of the more general theory of bulk and edge
excitations together.

\subsection{Edge states of the Pfaffian state}

The Pfaffian state \cite{mr}, for even particle number $N$, is defined by the
wavefunction           
\begin{equation}
 \Psi_{\rm Pf}(z_{1},\cdots,z_{N})={\rm Pf}\,\left(\frac{1}{z_{i}-z_{j}}\right)
\prod_{i<j}(z_{i}-z_{j})^{q}\exp[-\frac{1}{4}\sum|z_{i}|^{2}]
\end{equation}
where the Pfaffian is defined by
\[{\rm Pf}\, M_{ij}=\frac{1}{2^{N/2}(N/2)!}
\sum_{\sigma\in S_{N}} {\rm sgn}\, \sigma \prod_{k=1}^{N/2} M_{\sigma(2k-1) 
\sigma(2k)}\]
for an $N\times N$ antisymmetric matrix whose elements are $M_{ij}$. It 
is the lowest angular momentum ground state of the Hamiltonian \cite{gww}
\begin{equation}
H=V \sum_{<ijk>}\delta^{2}(z_{i}-z_{j})\delta^{2}(z_{i}-z_{k})
\label{pfaff3bodH}
\end{equation}
(where the sum is over distinct triples of particles) for the case $q=1$  and
of similar three-body short-range interactions for $q>1$. The filling factor is
$1/q$. The Pfaffian state is totally antisymmetric for $q$ even, so could 
describe electrons, while for $q$ odd it describes charged bosons in a 
high magnetic field.
Zero energy quasihole excitations correspond to increasing the flux inside the
area spanned by the fluid, as usual, but in this case the basic objects contain
a half flux quantum each and must be created in pairs. A wavefunction for two
quasiholes was proposed in \cite{mr}; it is 
\begin{eqnarray}
\lefteqn{\Psi(z_{1},\ldots,z_{N};w_{1},w_{2})=} \nonumber \\
 & & \frac{1}{2^{N/2}(N/2)!}\sum_{\sigma\in S_{N}} {\rm sgn}\, \sigma 
\frac
{\prod_{k=1}^{N/2} [(z_{\sigma(2k-1)}-w_{1})(z_{\sigma(2k)}-w_{2})+(w_{1}
\leftrightarrow w_{2})] }
{ (z_{\sigma(1)}-z_{\sigma(2)})\cdots(z_{\sigma(N-1)}-z_{\sigma(N)}) }
 \nonumber \\
 & & \qquad\times\prod_{i<j}(z_i-z_j)^q
\exp[-\frac{1}{4}\sum |z_{i}|^{2}].
\label{pfaff2qholes}
\end{eqnarray}              
It is clear that for $q=1$ the quasihole states are zero energy eigenstates of
$H$, (\ref{pfaff3bodH});
this also holds for the appropriate 3 body $H$ for $q>1$.
It will be seen that it is the pairing structure built into the ground state
which allows insertion of Laughlin-like factors which act on only one member of
each pair and hence effectively contain a half flux quantum each, unlike the
usual Laughlin quasihole that corresponds to a full flux quantum. The same
structure requires that quasiholes are made in pairs, since the wave function
must be homogeneous. 
When quasiholes coincide, that is, when $w_1=w_2$, a Laughlin quasihole is
recovered. 

The multi-quasihole states can be used to generate
the edge spectrum of the Pfaffian state. We initially used such an approach,
but now find it simpler to write down an ansatz which, we believe, 
in fact describes all the zero-energy
states. We construct wavefunctions for $N$ electrons ($N$ odd or even) which we
will interpret as having $F$ fermions created at the edge:
\begin{eqnarray}
\lefteqn{\Psi_{n_1,\ldots,n_F}(z_1,\ldots,z_N)=}\nonumber\\
& & \frac{1}{2^{(N-F)/2}(N-F)/2!}\sum_{\sigma\in S_{N}}{\rm sgn}\,\sigma 
\frac
{\prod_{k=1}^F z_{\sigma(k)}^{n_k}}
{ (z_{\sigma(F+1)}-z_{\sigma(F+2)})\cdots(z_{\sigma(N-1)}-z_{\sigma(N)}) }
\nonumber \\
 & & \qquad\times\prod_{i<j}(z_i-z_j)^q\exp[-\frac{1}{4}\sum |z_i|^2].
\label{ansatzpfedge}
\end{eqnarray}              
Here $0\leq n_1<n_2<\cdots <n_F$, are a set of distinct nonnegative integers,
and $N-F$ clearly must be even. The sum over permutations can be divided into a
sum of terms in each of which the unpaired electrons (those with indices
$\sigma(1)$ through $\sigma(F)$ in the above expression) are antisymmetrized
among themselves. Then each term contains a Slater determinant in these
coordinates, representing fermions in wavefunctions $z^{n_k}$, hence the stated
conditions on the $n_k$'s. 
The angular momentum of the states is
\begin{equation}
M=\sum_{k=1}^Fn_k + \frac{1}{2}(qN(N-1)-(N-F)).
\end{equation}
Hence the angular momentum relative to the ground state, $\Delta M= M-M_0$, is
\begin{equation}
\Delta M=\sum_{k=1}^F(n_k+\frac{1}{2}).
\label{deltaMpf}
\end{equation}
Note that the angular momentum of the ground state is calculated 
{\em at the same number of electrons}, and is $M_0=N(q(N-1)-1)/2$. Such a 
ground state only exists for $N$ even, but we use the formula as an 
interpolation for $N$ odd also, to yield (\ref{deltaMpf}).
We interpret the states as having fermions created in orbitals of angular
momentum $\Delta M=n+1/2$, $n=0$, $1$, \ldots, which is exactly the
description of the ground state (antiperiodic) sector for Majorana-Weyl
fermions on a circle. We should note, however, that if we choose a certain even
number $N_0$ of electrons and use the ground state for $N=N_0$ as a reference 
ground state, then the odd fermion number sectors occur only when some charge
has been either added or removed from the ground state. More precisely, 
odd fermion numbers can occur when $N$ is odd, so an odd amount of charge has
been added to the edge, and similarly for even numbers. Thus the parity of the
fermion number is equal to the parity of the (integral) charge added. 
This seemingly trivial observation indicates that the fermion and charge edge 
degrees of freedom are not completely decoupled. This is analogous to the 
spin-charge separation at the edge of the Halperin states, where although the 
spin and charge form separate excitations that can be moved along the edge 
independently, there are global selection rules that relate the total spin and 
charge added at the edge, in a similar way as here \cite{bf}. This is closely 
related to the absence of any spin-charge separation in the quasiparticles in 
the bulk, which can carry spin $1/2$ only if they also carry nonzero charge. 
These ``projection rules'' will be discussed more extensively later, including 
the hierarchy states, to one of which the Halperin state is isomorphic.

In addition to these states, we can also take any
one of them multiplied by a symmetric polynomial in all the $z_i$s, which is
again a zero energy eigenstate. These polynomials represent the ubiquitous
chiral bosons associated with charge excitations and need not be considered
further at the moment. To ensure that all these states represent linearly
independent edge excitations, we must certainly take the limit
$N\rightarrow\infty$ when studying each space of angular momentum eigenstates 
at $\Delta M$ fixed and finite. As we will see below, the states without
symmetric polynomial factors appear to be linearly independent, but a full
proof of this, and of independence of the symmetric polynomials, appears
difficult. In Appendix A we prove that all zero-energy states can be 
written as linear combinations of the forms (\ref{ansatzpfedge}) times 
symmetric polynomials. In Appendix B we indicate how we showed, for 
$\Delta M$ up to $8$, that all these states are linearly independent. 
This provides rather convincing evidence for our simple form 
(\ref{ansatzpfedge}).

For completeness, we also give expressions for the edge states in the other,
``twisted'',
sector, where the Majorana-Weyl field obeys periodic boundary conditions. These
states occur when an odd number of quasiparticles are present far inside the 
edge. For simplicity, we consider a single quasihole at the center of the disk.
The ground state in this sector can be produced by taking 
the two-quasihole state above, dividing by $w_2^{N/2}$ and letting 
$w_1\rightarrow 0$, $w_2\rightarrow\infty$. On including unpaired electrons as
above we obtain
\begin{eqnarray}
\lefteqn{\Psi_{n_1,\ldots,n_F}(z_1,\ldots,z_N)=}\nonumber\\
& & \frac{1}{2^{(N-F)/2}(N-F)/2!}\sum_{\sigma\in S_{N}}{\rm sgn}\,\sigma 
\frac
{\prod_{k=1}^F z_{\sigma(k)}^{n_k}
\prod_{l=1}^{(N-F)/2}(z_{\sigma(F+2l-1)}+z_{\sigma(F+2l)})}
{ (z_{\sigma(F+1)}-z_{\sigma(F+2)})\cdots(z_{\sigma(N-1)}-z_{\sigma(N)}) }
\nonumber \\
 & & \qquad\times\prod_{i<j}(z_i-z_j)^q\exp[-\frac{1}{4}\sum |z_i|^2].
\label{ansatzpfedge+qh}
\end{eqnarray}              
These states have angular momentum 
\begin{equation}
M=\sum_{k=1}^Fn_k +\frac{1}{2}qN(N-1).
\end{equation}
Hence the ground state, in which $F=0$, has angular momentum $M_0=qN(N-1)/2$ 
and the angular momentum relative to the ground state, $\Delta M= M-M_0$, is
\begin{equation}
\Delta M=\sum_{k=1}^Fn_k.
\label{deltaMpf+qh}
\end{equation}
Similar remarks to those above about the $F$ odd cases apply here. These quantum
numbers are exactly those expected for Majorana-Weyl fermions obeying periodic
boundary conditions, in which fermions can be added in orbitals with angular
momentum $n=0$, $1$, \ldots. In particular, note that
for $F=1$, we can take $n_1=0$ and obtain a state with zero increased angular
momentum. This state is entirely meaningful; since it has an odd number of 
electrons, it is not the same as the ground
state (with the added quasihole), which has an even number of electrons. 
That all states in this sector have the form shown is also proved in Appendix
A; proof of linear independence along the lines of Appendix B has been done at
low $\Delta M$.

Other numbers of quasiholes, either even or odd, are obtained by taking states
in the untwisted or twisted sectors, respectively, and multiplying in  
Laughlin quasihole factors $\prod_i z_i^m$. These states are
similar to those in the two sectors above but with additional charge at the 
edge, to which it has been pushed by the quasiholes at the
center. 

All these states with quasiholes at the center of the disk obey projection
rules similar to those for the states without the quasiholes. In this case, as 
$N$ varies, the charge at the edge runs over values which are integers plus a
fixed fraction (defined mod 1, and equal to a multiple of $1/2q$). 
If the charge added at the edge is defined relative to a reference ground state
at even $N=N_0$, for each number of quasiholes at the center, then the
projection rules are unchanged. This avoids problems of definition in the case
where $q$ Laughlin quasiholes have been added at the center. A more
satisfactory approach will be given for the case of two edges, as on the
cylinder, in \S IV. The states with quasiholes added at the center of the disk 
can be viewed as a special case of this. One could argue that for the disk, it
is more natural to exclude any bulk excitations, in which case there are no
twisted or quasihole sectors. The other sectors arise only when both edges are
present, or in the presence of bulk quasiparticles.

For completeness, we include a table of the dimensions of the spaces of fermion
states for low $\Delta M$ in the untwisted sector:
\begin{center}
\begin{tabular}{c|c|c|c|c|c|c|c|c}
	$\Delta M $  & 1 & 2 & 3 & 4 & 5 & 6 & 7 & 8   \\ \hline
	dim          & 0 & 1 & 1 & 2 & 2 & 3 & 3 & 5 
\end{tabular}\quad.
\end{center}
The entries in the table are the ones that have been verified in Appendix B.
The dimensions of the full space of edge excitations in the untwisted, even
particle number sector are
found by convoluting these numbers with those for the U(1) chiral boson system
given earlier. Thus all states in this sector have been verified to be
linearly independent up to $\Delta M=8$.
In a numerical calculation in \cite{wen3} only the states for
$\Delta M<5$, which in fact contain a maximum of two excited fermions, 
were found. The first state with 4 fermions excited appears at $\Delta M=8$ 
(and is included in the table).

\subsection{Edge states of the Haldane-Rezayi state}

The Haldane-Rezayi (HR) state \cite{hr} can be written in terms of the 
coordinates of $N/2$ up-spin electrons at $z_1^\uparrow$, \ldots, and $N/2$ 
down-spin electrons at $z_1^\downarrow$, \ldots as   
\begin{eqnarray}      
\lefteqn{\Psi_{HR}(z_{1}^{\uparrow},\ldots,z_{N/2}^{\uparrow},z_1^{\downarrow},
                       \ldots,z_{N/2}^{\downarrow})=}\nonumber \\
   & & \sum_{\sigma\in S_{N/2}}{\rm sgn}\,\sigma
         \frac{1}{(z_{1}^{\uparrow}-z_{\sigma(1)}^{\downarrow})^{2}
           \cdots(z_{N/2}^{\uparrow}-z_{\sigma(N/2)}^{\downarrow})^{2}}
\prod_{i<j}(z_{i}-z_{j})^{q}
         \exp[-\frac{1}{4}\sum|z_{i}|^{2}].
\label{HRground}
\end{eqnarray}           
Here $q$ is even to describe fermionic electrons, and the filling factor is
$1/q$. The first factor is of course just a determinant. The product over
$z_i$'s with no spin labels attached is over all particles.
The fact that this describes a singlet is discussed carefully in \cite{hr}.
Strictly speaking, the form given is an abuse of notation. The correct way to
write the functions is as a function of $N$ electron coordinates, numbered from
$1$ to $N$, half of which have up spin and half down, and the permutations are
over the subset of electrons of each spin. On including the proper sign
factors, the spatial wavefunctions can be combined with the spinor
wavefunctions of the $N$ electrons, and then summed over all ways of choosing
which electrons have which spin. In this way, wavefunctions that are
totally antisymmetric under exchange of both the space and spin labels of
particles are constructed. This procedure is standard and has been
described in the literature \cite{hr,yosh,hamermesh}; it can be used to
produce states of definite total spin. Since the construction 
of such states from the functions given below is straightforward, 
if tedious, it will be omitted, and we will continue to use the abused 
notation as in (\ref{HRground}).
In \cite{mr} it was pointed out that this state can be regarded as a BCS-type 
condensate of spin-singlet pairs of spin-1/2 neutral fermions that consist of 
an electron and $q$ vortices, from which the spin-singlet property can be more
easily understood. 
The HR state is the unique zero energy state at $N_\phi=q(N-1)-2$ flux of a
``hollow-core'' pseudopotential Hamiltonian that gives any two particles a
nonzero energy when their relative angular momentum is exactly $q-1$ \cite{hr}.

As for the Pfaffian state, excitations of
$hc/2e$ flux are expected and by flux quantization they
should occur in pairs. In exact analogy with the Pfaffian state, the 
wavefunction for two quasiholes is:
\begin{eqnarray}
\lefteqn{\Psi_{HR}(z_{1}^{\uparrow},\ldots,z_{N/2}^{\downarrow};
w_{1}, w_{2})=}\nonumber \\
  & & \sum_{\sigma\in S_{N/2}}{\rm sgn}\,\sigma
            \frac{\prod_{k=1}^{N/2} 
             [(z_{k}^{\uparrow}-w_{1})(z_{\sigma(k)}^{\downarrow}-w_{2})
                          +(w_{1}\leftrightarrow w_{2})] }
   {(z_{1}^{\uparrow}-z_{\sigma(1)}^{\downarrow})^{2}\cdots
    (z_{N/2}^{\uparrow}-z_{\sigma(N/2)}^{\downarrow})^{2}} 
\prod_{i<j}(z_{i}-z_{j})^{q}
           \exp[-\frac{1}{4}\sum|z_{i}|^{2}].
\end{eqnarray}           
Due to the spin-independence of the newly inserted factors acting on each pair
inside the sum over permutations, the 
state is still a spin-singlet and this suggests that the quasiholes carry no
spin. We will see further evidence of this later.
The two quasihole state is again a zero energy eigenstate of the ``hollow
core'' Hamiltonian where all pseudopotentials (including $V_{0}$) are
zero except $V_{q-1}$, {\it i.e.}, for two particles to interact their relative 
angular
momentum should be $q-1$. To see this fact, expand the inserted factors for each
pair in terms of powers of $z_k^\uparrow \pm z_{\sigma(k)}^\downarrow$. 
Due to the symmetry between $z_k^\uparrow$ and $z_{\sigma(k)}^\downarrow$ in
each factor, it is easy to see that
$z_k^\uparrow-z_{\sigma(k)}^\downarrow$  
must occur to an even power. Thus in the complete wavefunction, the absence of 
$(z_k^\uparrow-z_{\sigma(l)}^\downarrow)^{q-1}$ for any $k$, $l$, and hence the
zero energy property of the ground state, is preserved in the quasihole states. 

It is possible to write down directly the forms of all the zero energy states
of the hollow core Hamiltonian, in analogy with those for the Pfaffian. In the
untwisted sector, in terms of the coordinates of $N_\uparrow$ up electrons,
$N_\downarrow$ down electrons, the wavefunctions are linear combinations of
(omitting symmetric polynomial prefactors)
\begin{eqnarray}
& & 
\frac{1}{(N_\uparrow-F_\uparrow)!}
\sum_{\stackrel{\scriptstyle\sigma\in S_{N_\uparrow}}{\rho\in
S_{N_\downarrow}}}
{\rm sgn}\,\sigma \,{\rm sgn}\,\rho
\frac
{\prod_{k=1}^{F_\uparrow} (z_{\sigma(k)}^\uparrow)^{n_k}
 \prod_{l=1}^{F_\downarrow} (z_{\rho(l)}^\downarrow)^{m_l}}
{ (z_{\sigma(F_\uparrow+1)}^\uparrow-z_{\rho(F_\downarrow+1)}^\downarrow)^2
\cdots(z_{\sigma(N_\uparrow)}^\uparrow-z_{\rho(N_\downarrow)}^\downarrow)^2 }
 \nonumber \\
& & \qquad\times\prod_{i<j}(z_i-z_j)^q\exp[-\frac{1}{4}\sum |z_i|^2].
\label{ansatzhredge}
\end{eqnarray}              
Here $N_\uparrow-F_\uparrow=N_\downarrow-F_\downarrow$ is the number of
unbroken pairs, and we may assume the $n_k$'s, $m_k$'s are strictly increasing
as for those in the Pfaffian edge states. These functions have a structure
similar to the real space wavefunctions for a BCS state with some broken pairs,
that is with BCS quasiparticles added; the latter would have a similar form for
the sum over permutations, but the factors $z_i^{\uparrow n_k}$ would be
replaced by plane waves. Here, of course, they represent edge states, in which
the fermions do behave as if they occupied plane waves running along the
one-dimensional edge. As written, these states do not have definite spin, but
eigenstates of ${\bf S}^2$ and of $S_z$ can be constructed as indicated above.
Since the paired electrons form singlets, the spin is determined by the spin
1/2 unpaired fermions in the sums over $\sigma$ and $\rho$, which behave
identically to ordinary spin 1/2 fermions. Hence the possible spin states are
determined by adding the spins of electrons in different orbitals (labelled by
$n_k$ or $m_k$), with the only constraint that an orbital occupied with both an
up and a down fermion must form a singlet. 

The angular momentum of the wavefunctions given is
\begin{equation}
M=\sum_{k=1}^{F_\uparrow}n_k + \sum_{k=1}^{F_\downarrow}m_k
+\frac{1}{2}(qN(N-1)-2(N-F_\uparrow-F_\downarrow)).
\end{equation}
Hence the angular momentum relative to the ground state, $\Delta M= M-M_0$, is
\begin{equation}
\Delta M=\sum_{k=1}^{F_\uparrow}(n_k+1)+\sum_{k=1}^{F_\downarrow}(m_k+1).
\label{deltaMhr}
\end{equation}
Conformal invariance ideas suggest that this implies that the edge excitations 
are fermions of conformal weight 1,
not conformal weight 1/2 as for the Pfaffian state (\S IIB) and the 331 state 
(\S IID). This will be discussed further in \S III.
We note that the projection rule arising from our states is the same as for the
Pfaffian, i.e.\ the parities of the fermion and charge numbers are the same.
             
{}From these wavefunctions we can obtain the total numbers of the low-lying 
edge 
excitations (excluding symmetric polynomials) of the HR state in the untwisted 
sector at fixed even $N$:
\begin{center}
\begin{tabular}{c|c|c|c|c|c|c}
 $\Delta M$ & 1 & 2 & 3 & 4 & 5 & 6  \\ \hline
 dim        & 0 & 1 & 4 & 5 & 8 & 10 
\end{tabular}\quad.
\end{center}
Again, though the count of states based on the wavefunctions given could of
course be continued, the table has been terminated at the largest $\Delta M$ 
where we were able to verify linear independence directly (see Appendix B).
Note that spins larger than $1$ do not occur at these low $\Delta M$. Our 
numbers for spin $0$ and $1$ excitations agree, when convoluted with the U(1) 
numbers, with those calculated in Wen, Wu and Hatsugai \cite{wwh}, though our 
results for $N$ large extend to higher $\Delta M$ than they can attain in a 
small system without encountering finite size effects.

The twisted sector is again obtained by including a spin-independent quasihole 
factor, this time
$\prod_{l=1}^{N_\uparrow-F_\uparrow}(z_{\sigma(F_\uparrow+l)}^\uparrow+
z_{\rho(F_\downarrow+l)}^\downarrow)$, in the sum on permutations. The angular
momentum of the excited states relative to the ground state in this sector is
\begin{equation}
\Delta M=\sum_{k=1}^{F_\uparrow}(n_k+1/2)+\sum_{k=1}^{F_\downarrow}(m_k+1/2).
\label{deltaMhrtw}
\end{equation}

Finally, it is once again possible to multiply in factors $\prod_i z_i^m$ that 
add charge to the edge, which are spin independent and identical to those for 
the Laughlin states.

\subsection{Edge states of the 331 state}

The 331 state is just one of a family of two-component states, the so-called
$mm'n$ states, first introduced by Halperin \cite{halp84}. 
Using notation $\uparrow$, $\downarrow$ for the two components, even though 
they need not represent spin, and bearing in mind that similar remarks to those
at the beginning of \S IIC about constructing totally antisymmetric
wavefunctions apply here also, these states can be written:
\begin{equation}
\Psi_{mm'n}(z_1^\uparrow,\ldots,z_{N/2}^\downarrow)=
\prod_{i<j} (z_i^\uparrow-z_j^\uparrow)^m 
\prod_{k<l} (z_k^\downarrow-z_l^\downarrow)^{m'} 
\prod_{rs} (z_r^\uparrow-z_s^\downarrow)^n \exp [-\frac{1}{4}\sum_i |z_i|^2].
\label{mmnground}
\end{equation}
The general $mm'n$ state is the unique lowest total-angular-momentum ground 
state of a spin-dependent pseudopotential Hamiltonian, that generalizes
(\ref{pseudH}) 
to the two-component case, which gives positive energy to any state in which 
two $\uparrow$ or $\downarrow$ particles have relative angular momentum less 
than $m$ or $m'$, respectively, or in which an $\uparrow$ and a $\downarrow$
particle have relative angular momentum less than $n$.

For the case when the exponents in these states are of the form
$m=m'=q+1$, $n=q-1$, $q\geq1$, (which give filling factor $\nu=1/q$, and the
partial filling factors for $\uparrow$, $\downarrow$ are both $1/2q$; for 
brevity, we will continue to refer to this class of states with general
$q$ as the 331 state), 
then use of the Cauchy determinant identity
\begin{equation}
\prod_{i<j} (z_i^\uparrow-z_j^\uparrow)
\prod_{k<l} (z_k^\downarrow-z_l^\downarrow)
\prod_{rs} (z_r^\uparrow-z_s^\downarrow)^{-1} 
 = \det \left( \frac{1}{z_i^\uparrow-z_j^\downarrow}\right)
\label{cauchident}
\end{equation}
allows the ground states to be written in a paired form, similar to the 
Pfaffian and HR states \cite{halpnewport,ho}. This identity can be understood 
physically, in terms of the description of bulk fractional quantum Hall effect
wavefunctions as conformal field theory correlators \cite{mr}, as expressing 
bosonization of correlators of a chiral Dirac (or Weyl) field (on the right 
hand side) in terms of correlators in a Coulomb gas (or exponentials of a 
chiral scalar Bose field) (on the left hand side). In terms of BCS-type
pairing, this function describes $p$-type spin-triplet pairing, with each 
pair in the $S_z=0$ state of a spin triplet \cite{halpnewport,ho}.

We will extend the fermionized description immediately to include the 
edge excitations in the untwisted sector, omitting symmetric polynomial 
prefactors:
\begin{eqnarray}
& & \frac{1}{(N_\uparrow-F_\uparrow)!}
\sum_{\stackrel{\scriptstyle\sigma\in S_{N_\uparrow}}{\rho\in
S_{N_\downarrow}}}
{\rm sgn}\,\sigma \,{\rm sgn}\,\rho
\frac
{\prod_{k=1}^{F_\uparrow} (z_{\sigma(k)}^\uparrow)^{n_k}
 \prod_{l=1}^{F_\downarrow} (z_{\rho(l)}^\downarrow)^{m_l}}
{ (z_{\sigma(F_\uparrow+1)}^\uparrow-z_{\rho(F_\downarrow+1)}^\downarrow)
\cdots(z_{\sigma(N_\uparrow)}^\uparrow-z_{\rho(N_\downarrow)}^\downarrow) }
  \nonumber \\
 & & \qquad\times\prod_{i<j}(z_i-z_j)^q\exp[-\frac{1}{4}\sum |z_i|^2]
\label{ansatz331edge}
\end{eqnarray}              
which is particularly similar to the HR case. 
For the angular momentum we obtain
\begin{equation}
M=\sum_{k=1}^{F_\uparrow}n_k + \sum_{k=1}^{F_\downarrow}m_k
+\frac{1}{2}(qN(N-1)-(N-F_\uparrow-F_\downarrow)).
\end{equation}
Hence the angular momentum relative to the ground state, $\Delta M= M-M_0$, is
\begin{equation}
\Delta M=\sum_{k=1}^{F_\uparrow}(n_k+1/2)+\sum_{k=1}^{F_\downarrow}(m_k+1/2).
\label{deltaM331}
\end{equation}
This is the correct behavior for the states of a chiral Dirac (or Weyl) field, 
where the two types $\uparrow$ and $\downarrow$ denote particles and 
antiparticles. This is as expected from the general arguments based on the 
form of the bulk ground state wavefunction \cite{mr}, which, as we have 
mentioned above, includes a correlator of this same type of fields. The
projection rule is once again the same as for the Pfaffian.

The edge states can be reexpressed in bosonic form as 
\begin{equation}
{\cal F}^\uparrow{\cal F}^\downarrow \Psi_{q+1,q+1,q-1}
\end{equation}
in which ${\cal F}^\uparrow$ (${\cal F}^\downarrow$) are symmetric polynomials
in the $\uparrow$ ($\downarrow$) coordinates only, and the numbers
$N_\uparrow$, $N_\downarrow$ of $\uparrow$ and $\downarrow$ particles need not
be equal. 

This system also has a twisted sector obtained in a similar way as in the other
examples, by multiplying by factors $\prod z_i^\uparrow$ or 
$\prod z_i^\downarrow$ that represent the elementary quasiholes located at the
center of the drop. For a single such factor, this leads to a formula for 
$\Delta M$ like (\ref{deltaM331}) but in which the $1/2$'s in the expression 
are dropped. 

In the bosonic form, it is easy to see that all these states, both
twisted and untwisted, are zero-energy eigenstates for the above-mentioned 
pseudopotential Hamiltonian, as an extension of the arguments
for the one-component Laughlin states, and that they span the space
of such states. The equivalence of the bosonic and
fermionic forms of edge state wavefunctions involves generalizations of the
Cauchy determinant identity (\ref{cauchident}). We will return to the 
bosonized description in \S IVD.

\section{Field Theory of the HR State}
\subsection{Field theory of the edge states of the HR state}

We have seen in the previous section that, apart from the charge fluctuation 
excitations, the
edge excitations in the states we have studied are free fermions (for most of
this section, we ignore the projection rules exhibited in \S II; they will be
reincorporated in \S IV). For the
Pfaffian and 331 states, these can clearly be described by relativistic Fermi
fields in $1+1$ dimensions (i.e.\ distance along the edge and time) of scaling 
dimension 1/2, which we will denote $\psi$ for the Majorana field in the
Pfaffian case, and $\psi_\uparrow$, $\psi_\downarrow$ for the Dirac field and
its adjoint in the 331 case. These standard field theories need not be
described here. For the HR state, a natural candidate might
have been the Dirac theory, with up and down excitations described by particle 
and antiparticle. However, in the Dirac field theory, there is no SU(2) 
symmetry that can be generated by local expressions for the spin density and
current, and in fact it correctly describes the edge of the 
331 state. Moreover the angular momentum quantum numbers show that the
field for the HR state does not have scaling dimension 1/2, but instead
dimension 1. This puzzle will be addressed in this section. Another attempt at
its resolution has been made by Wen and Wu \cite{ww}, which described the bulk
wavefunctions, but did not exhibit the simple Lagrangian description shown 
here. 

First we write a Hamiltonian that reproduces the angular momentum 
eigenvalues already found. Introducing a velocity $v_s$ for
the spin excitations when a term $\lambda M$ is present, the Hamiltonian for an
edge of circumference $L$ would 
be
\begin{equation}
H=v_s\sum_{n=1}^{\infty} k (a_{k \uparrow}^{\dagger}a_{k \uparrow}+
a_{k \downarrow}^{\dagger}a_{k \downarrow}).
\label{HRedgeham}
\end{equation}
Here the operators $a_{k\sigma}$, $a_{k\sigma}^\dagger$, with $k=2\pi n/L$, 
obey the canonical anticommutation relations
$\{a_{k\sigma}^{\dagger}, a_{k'\sigma'}\}=\delta_{kk'}\delta_{\sigma\sigma'}$, 
$\{a_{k\sigma},a_{k'\sigma'}\}=\{a_{k\sigma}^\dagger,a_{k'\sigma'}^\dagger\}=
0$. Comparing with the result for the Pfaffian state, where there is a real
(Majorana), right-moving (Weyl) fermion, (see for example \cite{wen}) we see
that apart from the extra $S^z$ quantum number, the boundary condition is
periodic in the ground state (untwisted) sector here, while it was 
antiperiodic in the Majorana-Weyl system. 
 
Therefore we propose a new $1+1$ dimensional fermion theory for the neutral 
part of the HR edge, with a doublet of complex Fermi fields 
$\Psi_{\sigma}(x, t)$, $\sigma$ = $\uparrow$ or $\downarrow$, and a chiral 
Lagrangian density (inspired by that for chiral scalar bosons \cite{stone}) of 
the explicitly SU(2) invariant form:
\begin{equation}
{\cal L}=\frac{1}{4}\varepsilon^{\sigma\sigma'}(\partial_{t}-v_{s}
\partial_{x})\Psi_{\sigma}\partial_{x}\Psi_{\sigma'}
+ h.c.\,. 
\end{equation}        
Here $\varepsilon^{\sigma\sigma'}=-\varepsilon^{\sigma'\sigma}$, 
$\varepsilon^{\uparrow\downarrow}=1$.  
The canonical procedure leads to the following, simpler looking Hamiltonian
\begin{equation}
H=\frac{1}{2}\int dx\, 
v_{s}(\partial_{x} \Psi_{\uparrow}
\partial_{x} \Psi_{\downarrow}+\partial_{x} \Psi_{\downarrow}^{\dagger}
\partial_{x} \Psi_{\uparrow}^{\dagger})
\label{canHRham}
\end{equation}         
together with canonical momenta $\Pi_\uparrow = \partial_x\Psi_\downarrow$,
$\Pi_\downarrow = -\partial_x\Psi_\uparrow$.
Using periodic boundary conditions, and going to Fourier modes, we see that for
the zero wavevector modes we obtain the first-class constraints that the
corresponding momenta vanish: $\Pi_\sigma(k=0)\equiv0$. The constraints can be
included by simply omitting the zero modes henceforth in this chiral theory.
Quantization using canonical anticommutation relations then leads to quantized 
fields:
\begin{eqnarray}
\Psi_{\uparrow}&=&\sum_{k>0} \frac{a_{k\uparrow}}{\sqrt{kL}}
\exp -ik(x+v_{s}t) + \sum_{k>0} \frac{a_{k\downarrow}^{\dagger}}{\sqrt{kL}}
\exp ik(x+v_{s}t)  \\
\Psi_{\downarrow}&=&\sum_{k>0} \frac{a_{k\downarrow}}{\sqrt{kL}}
\exp -ik(x+v_{s}t) - \sum_{k>0} \frac{a_{k\uparrow}^{\dagger}}{\sqrt{kL}}
\exp ik(x+v_{s}t)
\label{fieldops}
\end{eqnarray}                                 
and their adjoints where 
$\{a_{k\sigma}^{\dagger}, a_{k'\sigma'}\}=\delta_{kk'}\delta_{\sigma\sigma'}$, 
$\{a_{k\sigma},a_{k'\sigma'}\}=\{a_{k\sigma}^\dagger,a_{k'\sigma'}^\dagger\}=0$,
and $k=2\pi n/L$  with $n=1$, $2$, \ldots\ for a system of circumference $L$. 
It is assumed that the vacuum $|0\rangle$ obeys $a_{k\sigma}|0\rangle=0$. The 
normal ordered version of (\ref{canHRham}) then yields (\ref{HRedgeham}).

Because of the Fermi statistics chosen for the field, and the positive-definite
norm imposed, as usual, on the Hilbert space, this field theory is not Lorentz 
invariant, in
spite of the gapless linear spectrum. Consequently, it is not conformally 
invariant either. This may be surprising since we have become used to the edge 
theories being some chiral conformal system, but in fact, since we started 
with a nonrelativistic system of electrons in a high magnetic field, nothing 
guarantees that the edge must be Lorentz invariant, even when there is a 
linear dispersion relation for the excitations. Nonetheless, we will see that
there is a closely related conformal field theory.

Returning to the chiral theory, the SU(2) currents can be found using the 
standard Noether procedure:
\begin{equation}
{\cal J}_{0}^{a}={1\over2}\sum_{\sigma,\sigma'} \frac
{\partial {\cal L}}
{\partial (\partial_{t} \Psi_{\sigma})}
i \tau_{\sigma\sigma'}^{a} \Psi_{\sigma'}+ h. c.
\end{equation}                                
where we specify the Grassmann derivative 
$\delta {\cal L}/\delta(\partial_t \Psi_i)$ to be taken from the left, 
 $\tau^{a}$ ($a=1,2,3$) are Pauli matrices, and normal ordering is assumed.
Then the total spin operators for the edge are
$S^a=\int dx\, {\cal J}_{0}^{a}(x)$, and 
\begin{eqnarray}
S^{z}&=&\sum_{k}{1 \over 2} (a_{k\uparrow}^{\dagger}a_{k\uparrow}-
a_{k\downarrow}^{\dagger}a_{k\downarrow})\\
S^+&=&\sum_k a_{k \uparrow}^{\dagger}a_{k \downarrow}\\
S^-&=&\sum_k a_{k \downarrow}^{\dagger}a_{k \uparrow}
\end{eqnarray}
which are easily seen to satisfy the SU(2) commutation relations. However, 
unlike other FQHE systems, such as the Halperin state, where the edge theory 
is not only conformally invariant but also has a Kac-Moody current algebra as a
spectrum-generating algebra, here the currents do not form a Kac-Moody algebra,
and their correlation functions contain logarithmic factors.

We are ready to identify the operators describing addition or removal of 
electrons at the edge in our conjectured edge field theory. The operators 
$\partial \Psi_{\uparrow}e^{-i\sqrt{q}\varphi}$ and 
$\partial \Psi_{\downarrow}e^{-i\sqrt{q}\varphi}$ represent the 
electron annihilation field operators $\Psi_{\uparrow el}$ and 
$\Psi_{\downarrow el}$.
In these expressions $\varphi(x,t)$ is the usual chiral boson field 
representing the density fluctuations at the edge, with propagator
\begin{equation}
\langle \varphi(x,t)\varphi(0,0)\rangle=-\ln(x+v_c t);
\end{equation}
it is related to the 
electronic charge density by $\rho = -i\partial \varphi/\sqrt{q}$ 
\cite{wen2,mr}. The exponential of the boson operator creates a bosonic object 
whenever $q$ is even, so the scalar fermion, like the Majorana fermion in the 
Pfaffian case, field makes the whole thing into a fermion, as the 
electron should be. Note that the gradient of the scalar fermion field appears 
here, not the field itself; this reproduces the spin $1$ field found earlier.
Taking the charge excitations to propagate with velocity $v_c$,
we find that the electron propagator is
\begin{equation}
\langle 0|\Psi_{\uparrow el}^{\dagger}(x, t)\Psi_{\uparrow el}(0,0)|0\rangle 
\propto
\frac{1}{(x + v_s t)^2(x+v_c t)^q}
\end{equation}
where the space-time separation of the two fields should be small compared with
the circumference of the disk. The total exponent is thus $q+2$, in contrast to
that for the Pfaffian and $331$ state at $\nu=1/q$ which give $q+1$, whereas
the Laughlin states give $q$. Consequently, the expectation $n(k)$ of the 
occupation number of the $k$th single {\em electron} state, which can be 
obtained by Fourier transforming (in one dimension, along the edge) the 
equal-time electron Green's function, has a power-law singularity 
$n(k)\sim|k-k_{\rm max}|^{q+1}$ for the HR state, while the exponent is $q$ 
for the Pfaffian and $331$ states, $q-1$ for the Laughlin states (for the full
Landau level, $q=1$, and there is a discontinuity in $n(k)$).
Numerical simulations have been performed for both the Haldane-Rezayi and
Pfaffian states \cite{rezhaldedge}, but no conclusion about the exponents in 
occupation number $n(k)$ versus $k$ relevant to the edge field theory is 
drawn in the published work.

One might also expect that, just as the bulk system has pairing of the composite
fermions \cite{mr}, similar algebraic BCS-type expectations should appear 
at the edge. Indeed, because of the form of the operators (\ref{fieldops}), 
we have, for example,
\begin{equation}
\langle 0|\partial\Psi_\uparrow(x,t)\partial\Psi_\downarrow(0,0)|0\rangle 
\propto
\frac{1}{(x + v_s t)^{2}}
\label{dpsidpsicorr}
\end{equation}
which can be viewed as a pairing function. However, this correlator omits the
exponentials of $\varphi$ needed to represent the electron operators; if these
were included, the correlator would decay rapidly, since the fields carry the
same, not opposite, charge. This could be taken to illustrate, for the edge 
theory, how pairing occurs for the composite fermions, not for electrons. 
Similar phenomena can be found in the Pfaffian and 331 states. However, while 
the correlator shown is legitimate as it stands for the scalar fermion field 
theory, it is not a legitimate correlator for the HR edge theory, because the 
required intermediate states, where only a single fermion has been added or 
removed from the ground state (or excited states, in the finite temperature 
case), do not obey the projection rule found in \S IIC. Since fermions can be
created from the ground state only by breaking pairs, states with an odd number
of fermions occur only when an odd number of charges have also been added at
the edge (assuming there are no changes in the interior of the system). Only 
operators that respect this rule can be constructed in the edge field theory. 
Thus, in the theory of the HR edge, this correlator can be constructed only 
for equal times, and must then be viewed as the expectation of a single, 
nonlocal, operator.

\subsection{Conformal field theory of the bulk HR state}

In \cite{mr} a mathematical connection between FQH wave functions and
correlators in CFT was presented, namely the elegant wavefunctions of some
important FQHE states are actually correlators (or conformal blocks) in a
chiral 2-dimensional conformal field theory. However, the question of what
theory this would be for the HR state was left unresolved. The difficulty was
to understand the SU(2) symmetry (in fact, the singlet nature of the state)
in terms of CFT. We will present a solution to this problem here. The natural 
choice for the CFT in the HR case is a nonunitary theory with the Euclidian 
action (containing at this stage both right and left moving degrees of freedom 
for convenience):
\begin{equation}
S=\int \frac{d^{2}x}{8 \pi}\varepsilon^{\sigma\sigma'} 
\partial_{\mu} \Psi_{\sigma} \partial_{\mu}\Psi_{\sigma'}
\end{equation}        
and the Grassman field $\Psi_\sigma$ is regarded as real. Thus $\Psi_\sigma$ is
a relativistic scalar fermion and this model is conformally invariant, but its
states do not all have positive self-overlaps because of the violation of the
spin-statistics connection, so we say that the Virasoro
representations are nonunitary; the central charge is $c=-2$. 
(We note that the spin-statistics theorem relates the statistics of the fields
in a positive-definite, Lorentz-invariant field theory to the ``spin'' defined
by rotations of the Euclidian two-dimensional spacetime, which in a conformal
theory is the difference of the right- and left-moving conformal weights, and
not to what we have been calling spin, which describes the transformation
properties under SU(2) rotations that leave the spatial coordinates
unchanged.)
This system is an
anticommuting counterpart to the system of a pair of scalar boson fields which 
is unitary and has $c=2$. $\partial\Psi_\sigma$ is a field of conformal weight 
$1$ whose correlators
\begin{equation}
\langle\partial\Psi_\uparrow(z_1^\uparrow)\cdots\partial\Psi_\downarrow
(z_{N/2}^\downarrow)\rangle
\end{equation}
reproduce the determinant in the HR state. The action $S$ is manifestly 
invariant under the group of real symplectic transformations Sp(2,$\bf R$)
(which preserves the reality property of the fields, and has the same
complexified Lie algebra as SU(2)) and thus the correlator produces a singlet 
(since the vacuum is invariant). However the Noether currents are of the form
$\Psi_\sigma\partial\Psi_{\sigma'}-(\sigma\leftrightarrow\sigma')$ which are 
not ``good'' conformal fields
since their correlators contain logarithms, so there is no Kac-Moody symmetry.
We note that Wen and Wu \cite{ww} arrived at an equivalent description of this 
$c=-2$ CFT system in terms of OPEs but did not give the simple lagrangian
description above.

It is possible to construct ``twist'' fields \cite{dfms} for the field 
$\Psi_\sigma$, which
play a role similar to the spin field of the Ising (or Majorana) field theory
in the construction of the bulk quasihole wavefunctions \cite{mr}. These 
fields obey identical relations as those defined in the next subsection, so we
postpone discussion until then.

\subsection{Relation of bulk and edge field theories}

Next we will explain briefly how the relation of the bulk and edge field
theories can be used in order to define quasihole operators at the edge. 

The field theories as defined in \S\S IIIA, B appear very similar. The
difference is that, in \S IIIA the fields were not required to satisfy
a reality condition, they and their adjoints both appeared in the action,
and the Hilbert space was found to have positive norms (but no Lorentz
invariance), while in \S IIIB fields were real, Lorentz invariance was
maintained and the self-overlaps of some states were negative. Here we will
consider the positive definite theory of \S IIIA, and exhibit a conformal
structure in this system. This does not contradict the earlier statements
because the stress-energy tensor involved is not self-adjoint (with respect to
this inner product).

The correlators of the (gradients of the) fields in this system, as already 
exhibited in (\ref{dpsidpsicorr}), are clearly conformally invariant. If we work
in imaginary time $\tau$, and use the spacetime coordinate $z=x+iv_s\tau$, 
then the fields obey the operator product expansion (ope)
\begin{equation}
\partial\Psi_\sigma(z)\partial\Psi_\sigma'(0)
        \sim\varepsilon_{\sigma\sigma'}/z^2
\label{ope}
\end{equation}
up to the usual less singular terms (these have the same form as those for the
real fields in the theory in \S IIIB). Then, {\em if we consider only 
correlators of $\partial\Psi_\sigma$, not of $\partial\Psi^\dagger_\sigma$}, 
these correlations are conformally invariant. The stress-energy tensor that 
generates these transformations is 
$T(z)=-\varepsilon_{\sigma\sigma'}:\partial \Psi_{\sigma}
\partial\Psi_{\sigma'}:/2$, which can be verified to obey the ope's 
of a stress-energy tensor using only the ope (\ref{ope}). This operator is not
self-adjoint, so its Fourier components $L_n$ in general do not obey 
$L_{n}^{\dagger}=L_{-n}$. However, the Hamiltonian $L_0$ derived from $T$
coincides with that found above (\ref{HRedgeham}). Naturally, if we consider 
instead $\partial\Psi^\dagger_\sigma$, there will be another non-hermitian 
stress tensor generating conformal transformations of those correlators.      
The vacuum is annihilated by the modes $L_n$, $n\geq -1$ of either of these
stress tensors, as required in a conformal field theory.

We may now consider twist operators in this theory, which quite generally are
operators that twist the boundary conditions on the fields, in the manner
already described in \S II. We will require these to be also Virasoro primary 
conformal fields for the stress tensor $T$ above, as they are in the
non-positive definite theory. 
They are introduced by the operator product expansion
\begin{equation}
\partial \Psi_{\sigma}(z) {\cal S}(w) \sim
  \frac{1}{\sqrt{z-w}} \hat{\cal S}_{-\sigma}(w)   
\end{equation} 
where $\cal S$ is a twist field and $\hat{\cal S}_{\sigma}$ is an excited 
twist field of spin $\sigma$. In the presence of the $\cal S$ fields the 
correlator for $\partial \Psi_{\uparrow}$ and $\partial \Psi_{\downarrow}$
which, in the untwisted sector, is given by the leading term in the ope
(\ref{ope})
\begin{equation}
\langle\partial \Psi_{\uparrow}(z) \partial\Psi_{\downarrow}(w)\rangle
=\frac{1}{(z-w)^2},
\end{equation}
becomes                          
\begin{equation}
\langle{\cal S}(\infty)\Psi_{\uparrow}(z)\Psi_{\downarrow}(w){\cal S}(0)\rangle
=\frac{\frac{1}{2}(\sqrt{\frac{w}{z}}+\sqrt{\frac{z}{w}})}{(z-w)^{2}}, 
\end{equation}
and by a standard calculation \cite{dfms} we come to the 
equation
\begin{equation}
\langle {\cal S}(\infty)T(z){\cal S}(0)\rangle =- \frac{1}{8} \frac{1}{z^{2}}
\end{equation}
which means that the conformal weight for the primary field $\cal S$  is 
$-1/8$, and that the correlator
\begin{equation}
\langle {\cal S}(z){\cal S}(0)\rangle=z^{1/4}
\end{equation}
increases with separation.
Negative scaling dimensions cannot appear in a conformal field theory
described by unitary representations of the Virasoro algebra. Since, in 
theories obeying the BPZ axioms, operators correspond one-to-one with states, 
they
indicate the existence of states with energy below that of the ground state. 
Since our Virasoro generators $L_n$ generate a nonunitary representation, this
is not a problem here. Moreover, $L_0$ coincides with the physical Hamiltonian
in both the untwisted and twisted sectors, at least up to an overall constant
in the twisted case. (We note that in order to calculate this from $\Delta M$ 
of the zero energy states a term related to the contribution of the bulk to
the angular momentum must be subtracted; see the next section.) We now propose 
that this constant is $-2\pi v_s/8L$ as predicted by the conformal 
considerations above, since this appears to contradict no principles. There is 
no real inconsistency in asserting that the energy of the supposedly 
``excited'' twisted ground state lies below that of the untwisted ``true'' 
one. We could take the twisted sector (where $\Psi_\sigma$ obeys antiperiodic 
boundary conditions) as the ground state sector. We do not do so because 
(i) the theory for the disk (i.e.\ the chiral theory) clearly identifies the 
periodic sector as the ground state, which has no quasiholes in the bulk; 
(ii) the ground state in the antiperiodic sector is not invariant under 
${\rm SL}(2,{\bf C})$ generated by $L_0$, $L_{\pm1}$ as required in a 
conformal theory, whereas that in the periodic sector is.

For the edge states of the HR quantum Hall state (as in all the paired 
theories considered in this paper), the twisted sector of the
fermions occurs only when the charge added at the edge is $1/2q$ (modulo
$1/q$). Making use of
conformal arguments for the states with added charge, we expect the
contribution to the energy from the charge sector be
$(2\pi v_c/L)Q^2/2q$, where $Q/q$ is the charge added to the edge, as 
will be discussed in the next section. Thus, depending on the ratio $v_c/v_s$, 
the net energy of these states in the HR case will be positive in most cases, 
except for the lowest added charge sectors. In particular, for 
$\nu=1/2$, the orginal case of interest for HR, there will be a sector with 
ground state energy $-2\pi v/16L$ if $v_s=v_c$. These sectors correspond to
operators of the form ${\cal S}e^{i\varphi/2\sqrt{q}}$, which are also the 
operators used in the bulk conformal field theory to generate quasihole states 
that are single valued with respect to the electron operators \cite{mr}.

\section{Edge States and their field theories on a Cylinder}
\subsection{General results and the Laughlin states}

In this section we consider zero energy states on a cylinder. For the Laughlin 
states, considered in the present subsection, the structure of the edge states 
is well known (see especially \cite{wen}), but will be reviewed here to ensure 
that the ideas are clear, and so as to introduce the partition function for 
two oppositely-moving edges.

On a right cylinder of circumference $L$, we work in the Landau gauge.
In terms of a complex coordinate $z$, the single particle
wavefunctions in the lowest Landau level are $e^{2\pi i n z/L}
e^{-\frac{1}{2}y^2}$, where $n$ is an integer and $y={\rm Im}\, z$; it has been
assumed that the boundary condition is that wavefunctions are periodic under
$z\rightarrow z+L$. A more general boundary condition is that the wavefunction
changes by $e^{i\phi}$ under such a transformation; in that case the
wavefunctions become $e^{i(2\pi  n +\phi)z/L} e^{-\frac{1}{2}y^2}$. Returning
to the case $e^{i\phi}=0$ from here on, we can write many-particle 
wavefunctions in
terms of $Z_j=e^{2\pi iz_j/L}$, for example the Laughlin state \cite{thoul84}:
\begin{equation}
\Psi_{\rm L}=\prod_{i<j}(Z_i-Z_j)^q \exp[-\frac{1}{2}\sum_i y_i^2].
\label{laughcyl}
\end{equation}
As any $z_i$ approaches any $z_j$, this function clearly retains the
properties of the Laughlin state in the plane, namely it vanishes as the $q$th
power. 

The Landau gauge has explicit symmetry under translation around the
cylinder, and the corresponding conserved quantum number, $M$, is the sum of 
the powers $M_i$'s of the $Z_i$'s in the many-particle wavefunctions. $M$ is 
the angular momentum; alternatively we could use the linear momentum equal to 
$2\pi M/L$. For $e^{i\phi}=1$, $M$ is integral. In the Laughlin state above, 
$M=qN(N-1)/2$, and the $M_i$'s of individual particles are in the range $0$, 
\ldots $q(N-1)$. The
single particle wavefunction with $M_i=n$ is peaked at $y=-2\pi n/L$, so the
Laughlin state occupies a corresponding range in the $y$ direction. Due to
translational symmetry in the $y$ direction, there are, however, also an 
infinite number of other Laughlin states obtained by shifts, which are 
produced by acting repeatedly on the wavefunction with $\prod_i Z_i$ 
(or its inverse). This operator shifts the $M_i$ by $1$, so since the filling
factor is $1/q$, it corresponds to shifting a charge $1/q$ from one edge to the
other. Fractional shifts change the boundary condition on the wavefunction, so
are not allowed in the Hilbert space at fixed $e^{i\phi}$.

The infinite set of Laughlin states (``ground states'') are the most dense or 
compact zero energy states, in the sense that the $M_i$'s lie in a range of 
minimum possible width (note that the pseudopotential 
Hamiltonian in arbitrary geometry is defined in terms of the order of vanishing
of the wavefunctions as two particles approach each other, which in the plane 
is equivalent to relative angular momentum). The total angular 
momentum of the Laughlin states is $M=\frac{1}{2}qN(N-1)-pN$, for the state 
where we have applied $\prod Z_i^{-p}$, $p$ integral, to the Laughlin state 
(\ref{laughcyl}). The range of $M_i$ values found in this state is then 
$-p\leq M_i\leq q(N-1)-p$. (Note that for $q(N-1)$ even, we could choose $p$
such that $M=0$ and obtain a state symmetrical about $M_i=0$.) All other zero
energy states have a broader range of $M_i$'s, and are obtained as edge
excitations of the two edges. In the $N\rightarrow \infty$ limit, the 
different Laughlin states become infinitely far apart in angular momentum, and 
the assignment of edge excitations as belonging to a particular ground state 
(from which their angular momentum differs by a finite amount) becomes 
unambiguous. There are then two sets of elementary edge 
excitations, out of which these excited states can be built, and as for the
disk, these are linearly independent in the $N\rightarrow \infty$ limit. 
The elementary bosons that create them are the operators $s_n=\sum Z_i^n$ and 
$\overline{s}_n=\sum Z_i^{-n}$ ($n>0$). We refer to the excitations created by 
the action of the $s_n$'s, whose contribution to $\Delta M_{\rm tot}$, the 
change in $M$ relative to the corresponding Laughlin state, is positive, as 
right-moving. The other operators, $\overline{s}_n$'s, create excitations at 
the other edge, have $\Delta M_{\rm tot}<0$, and are viewed as left-moving. 
(Here right and left refer to the two directions parallel to the edge.) 
Thus, if we can split $\Delta M_{\rm tot}$ into $\Delta M-\overline{\Delta M}$,
where the two terms are the contributions at the two edges, then 
we would like to view $\Delta M+\overline{\Delta M}$ as the ``pseudoenergy'' 
(within a scale factor), where both terms are defined to be non-negative for 
the case in the present subsection. The direction of motion then follows from 
the group velocity, the change in pseudoenergy with momentum of an elementary 
excitation.

A more realistic, and well-defined, method would be to introduce a Hamiltonian 
that breaks the degeneracy of the zero-energy ground states and 
edge excitations; for example, a parabolic confinement potential 
$\sum_i M_i^2$, would suffice, but unfortunately the wavefunctions that can be 
easily written down are not eigenstates of such a form, except for $q=1$. 
We assume that as such a term is turned on and the eigenstates evolve, they 
stay in one-one correspondence with those found here. We expect on physical 
grounds that excitations that move electrons further from the minimum of this 
potential, as the edge excitations do, have higher energy, and so for small 
$|\Delta M|$ the bosons have a dispersion relation $E\sim |\Delta M|$ for 
either edge, and form the modes of a (non-chiral) scalar Bose field. The 
effect of the Hamiltonian on the low-lying states of the system can then be 
determined through a renormalization group analysis, as has been done for the 
present case already \cite{kanefish}. In fact, the analysis of the effective 
field theory of low-lying excitations, and its operator content, given here is 
the basis for such an analysis for the paired states.

If the electron number in the Laughlin ground state is changed by $1$, the 
width changes by $q$ units. We can view this as a new operation that adds
charge to a single edge, without disturbing the bulk ground state. The full 
set of possible ground state systems can then be parametrized by the 
positions of the
two edges, which are almost independent. Each edge can be shifted by $q$ units
without affecting the other. Shifts by $m=1$, \ldots, $q-1$ must be performed
on both edges together. A shift by $q$ units at both edges is equivalent to 
removing an electron from one edge and inserting it in the ground state at the
other. If we extend the idea of a charge sector to include in a single sector 
all those that differ by integral charges, then there are only $q$ sectors, of 
charge $0$, $1/q$, \ldots, $1-1/q$ (modulo integers), where the charge is 
shifted by the stated amount from one edge to the other. For the 
pseudopotential Hamiltonian without the confining potential, these $q$ sectors 
are on an equal footing and the choice of zero is arbitrary. Above these 
ground states, bosonic excitations can be created at either edge, and have the 
same character in all sectors.
Thus, the Hilbert space of the edge excitations can be written in the form
\begin{equation}
{\bf V}= \bigoplus_{r=0}^{q-1}{\bf V}_{r/q}\otimes \overline{\bf V}_{r/q} 
\end{equation}
Each Hilbert space ${\bf V}_{r/q}$ ($\overline{\bf V}_{r/q}$), 
$r=0$, $1$, \ldots $q-1$, is the span of the full set of states in the 
extended charge sectors at the right- (left-) moving edge. 
The conformal field theory of the edge states that includes the
operators of charge $1$ in the chiral algebra is known as the ``rational
torus'' (see, e.g., \cite{mr}). Loosely, the chiral algebra is the algebra of
operators that affect only a single edge. There is such an algebra for both
edges, the two algebras are isomorphic, and operators in one commute or 
anticommute with those in the other. 
Each Hilbert space ${\bf V}_{r/q}$ ($\overline{\bf V}_{r/q}$), $r=0$, $1$, 
\ldots $q-1$, is an irreducible representation of the
fully-extended right- (left-) moving chiral algebra.

To extend our definition of the pseudoenergy, which was given for the edge
excitations of a single Laughlin state at fixed (but large) particle number, 
to the full set of sectors just described, we introduce an arbitrary reference 
ground state with $N=N_0$ particles and shift $p=p_0$, so that the angular 
momentum $M_0=\frac{1}{2}qN_0(N_0-1)-p_0N_0$. We calculate $\Delta M_{\rm tot}$
for Laughlin states where an integral amount of charge has been added to a 
single edge, by changing $N$ and adjusting the shift $p$ from $p_0$ such that 
either the maximum or minimum occupied $M_i$ is unchanged compared with the 
reference state in the case of charge added to the left- or right-moving edge, 
respectively. After subtracting a quantity related to the bulk of the system, a
step analogous to measuring momentum from the Fermi wavevector in a Fermi gas, 
this gives a formula for $\Delta M$ or $\overline{\Delta M}$,
valid in these special cases. This will then be used in all the sectors.
The occupied states in general lie in the interval 
$-p\leq M_i\leq qN(N-1)/2-p$. To add charge at the right-moving edge, we let 
$N=N_0+\Delta N$, and $p=p_0$. Then we calculate
\begin{equation}
\Delta M_{\rm tot}=q\Delta N^2/2 +\Delta N [q(N_0-1/2)-p_0].
\end{equation}
Therefore if we define $E\equiv\Delta M_{\rm tot}-[q(N_0-1/2)-p_0]
\Delta N$ for the excitation pseudoenergy at
the right edge, we obtain $E=q\Delta N^2/2$, for the Laughlin states, and this
is consistent with result for the charge fluctuation bosons, which do not
change the charge at the edge. Similarly, for the left edge,
where we must use the adjusted shift $p=p_0+q\Delta N$ so that the right edge
is unmoved, we obtain
\begin{equation}
\Delta M_{\rm tot}=-q\Delta N^2/2+\Delta N[-q/2-p_0].
\end{equation}
Then for the excitation pseudoenergy $\overline{E}$ at this edge we must use 
$\overline{E}=-\Delta M-[q/2+p_0]\Delta N$. For each edge, the coefficient of 
$N$ is the mean of the angular momenta of the highest (respectively, lowest) 
occupied single particle states in the reference state with $N_0$ and with
$N_0+1$ electrons, so it resembles the Fermi wavevector in a Fermi sea, on two
sides of the Fermi sphere. We now use these formulas also for the ground 
(and excited) states in any charge sector, having a combination of electrons 
added to $N_0$, shifts of charge from one edge to the other, and charge
fluctuations; for such states we can always 
calculate, for each edge separately, the amount of charge effectively added, 
which may now be fractional, but is always a multiple of $1/q$. 

All information about the number and quantum numbers of the edge excitations
in the thermodynamic limit (taken with $L^2/N$ fixed) can be conveniently
summarized in a partition function analogous to that in \S IIA. The partition 
function is a double series in a complex parameter $x$ and its complex 
conjugate $\overline{x}$, which contains information about right and left 
movers, respectively. In fact, if $H$ is the Hamiltonian whose eigenvalues 
are $2\pi v/L$ times the sum of an $E$ and an $\overline{E}$ found above, 
where $v$ is the speed of propagation of the edge excitations, and taking 
$x=\overline{x}=e^{-2\pi\beta v/L}$, then this partition function is the Gibbs 
grand canonical partition function of statistical mechanics, 
${\rm Tr}e^{-\beta H}$. (This is consistent with the assignments in \S II and 
\S III.) The $E$'s of the charge sectors we have found agree 
with the conformal weights that are found in the conformal field theory
\cite{wen}. The result is expected to be the same for a more realistic
Hamiltonian with a confining potential, as discussed above. 
The partition function is defined as
\begin{equation}
{\cal Z}(x,\overline{x})={\rm Tr}\,x^{E}\overline{x}^{\overline{E}}.
\end{equation}
The structure of $\bf V$ given above now allows us to express $\cal Z$ in 
terms of the trace over each space ${\bf V}_{r/q}$, and we define (using the 
Euler partition sum $p(n)$ for the bosons from \S IIA)
\begin{equation}
\chi_{r/q}^\pm(x)\equiv\sum_{m=-\infty}^\infty (\pm 1)^m x^{(mq+r)^2/2q}
\prod_{n=1}^\infty(1-x^n)^{-1}
\end{equation}
and the complex conjugate for $\overline{\bf V}_{r/q}$. The traces, such as the 
$\chi_{r/q}^+$'s, over the right-
and left-moving spaces are known as characters, since
they are essentially the characters, in the algebraic sense, of the irreducible
representations ${\bf V}_{r/q}$ of the chiral algebra. (The greater generality
afforded by the insertion of $\pm1$, and by allowing $r$ to be an arbitrary
real number, will be useful later.) The partition 
function can then be written as 
\begin{equation}
{\cal Z}(x,\overline{x})=\sum_{r=0}^{q-1}|\chi_{r/q}^+(x)|^2.
\end{equation}
Similar structures to those found here for the Laughlin states on a cylinder 
will be found for the other states in the following subsections; unfortunately,
for the paired states, while the method for calculating $E$ and $\overline{E}$ 
produces a similar contribution for the different charge sectors, it is too 
crude to produce the analogous energies that we expect to originate in the 
fermion sectors, such as the $-1/8$ discussed in the preceding section, though 
we believe that they could in principle be obtained in a refined calculation.

\subsection{Pfaffian state}

As for the Laughlin states, the zero-energy states on the cylinder can be
obtained from those in \S IIB by replacing $z_i$ by $Z_i$, the gaussian factor
by $\exp[-\frac{1}{2}\sum_i y_i^2]$, and recalling that the exponents of the 
$Z_i$'s run over all the integers, positive and negative. The formula for 
$\Delta M$ of the edge excitations in the untwisted sector containing 
fermions only still applies,
\begin{equation}
\Delta M_{\rm tot} = \sum_{k=1}^F(n_k+1/2)
\end{equation}
but now $n_k\geq0$ describes right-moving fermions, $n_k<0$ left-moving. For
the latter we can define $\overline{\Delta M}=-\Delta M_{\rm tot}$. 

By inspection of the resulting states, we deduce that, in the untwisted sector,
if $N$ even is fixed at $N_0$, then the total 
number of fermions excited is even, and the parity of the number at each edge
(i.e.,\ whether it is even or odd) must be the same. But if we increase $N$ by
$1$, we must create or destroy a fermion at one edge, as well as increase the 
charge by $1$, which can be done at the same edge without affecting the other. 
So the chiral algebra includes the operator $\psi e^{i\sqrt{q}\varphi}$ which 
does this \cite{mr}. 
Then the parity of the fermion numbers at the two edges can be opposite 
provided the parity of the integral amount of charge added relative to the 
reference state is also opposite. Applying the operation, or its adjoint, once 
more to the same edge, we find states where the charge has changed by $0$ or 
$2$ relative to the reference state, but the number of fermions has the same 
parity, still without affecting the other edge. Thus all these states lie in
the same extended charge sector, and similar results hold in the other
untwisted sectors where a fractional charge $m/q$ has been shifted from one
edge to the other. A total of $2q$ sectors results from these considerations. 

The twisted sector is obtained from the untwisted by inserting a factor 
$$\prod_{l=1}^{(N-F)/2}(Z_{\sigma(F+2l-1)}+Z_{\sigma(F+2l)})$$ in the states 
with $F$ unpaired fermions (in the notation of \S IIB) 
to produce the effective shift by a half-unit that transfers
charge $1/2q$ from one edge to the other. The angular momentum of the excited
fermions is then
\begin{equation}
\Delta M_{\rm tot} = \sum_{k=1}^F n_k
\end{equation}
where once again $n_k$ can run over the integers. This time $n_k>0$ represents
right-movers, $n_k<0$ left-movers. $n_k=0$ is the zero mode, which
cannot be assigned to the right or left moving sectors. On the other hand, the
requirement that the {\em total} number of fermions must be even whenever the
charge added to the reference state is even, and odd when it is odd, still
applies. Thus we have the general projection rule that applies in all the
sectors, twisted and untwisted, that the total number of fermions created
in right-moving, left-moving or zero modes together must be equal to the parity
of the total charge added to the system at the two edges together.  
In the twisted sectors, it can be satisfied by allowing either parity of 
right- and of left-moving fermions in all charge sectors, then choosing the 
occupation number (either $0$ or $1$) of the zero mode to obey the condition. 
Consequently, the distinction between ``even'' and ``odd'' sectors, that 
existed among the untwisted sectors and was responsible for the factor $2$ in 
the $2q$ sectors, no longer applies, and there are just $q$ distinct sectors 
(or irreducible representations of the chiral algebra). The total number of 
sectors is therefore $3q$, which, in line with the general connection between 
bulk and edge states made in \cite{mr}, is the same as the number of zero 
energy ground states found in the toroidal geometry \cite{gww}. Note that in 
this case the description of the chiral algebra as affecting only a single 
edge is not quite correct, because the right-moving operator $\psi$, and its 
left-moving analogue $\overline{\psi}$, each contain a term that changes the 
occupation number of the zero mode. Nonetheless, in the Majorana field theory, 
these operators anticommute; similarly, the algebra of operators assigned to 
one edge does (anti-)commute with those assigned to the other, even in the 
twisted sector.

The calculation of the partition function, which formalizes the above remarks,
is conveniently performed in terms of characters. The
basic objects are characters for the states at one edge that differ in charge
only by integers, and, in the untwisted sector, the parity of the charge 
difference from that in the lowest energy state is equal to the parity of the 
change in fermion number. Characters for the fermions alone will be useful; 
these are, for untwisted (antiperiodic) boundary conditions,
\begin{eqnarray}
\chi_{0}^{\rm MW}(x)&=&
\frac{1}{2}\left[\prod_{n=0}^\infty(1+x^{n+1/2})+
                          \prod_{n=0}^\infty(1-x^{n+1/2})\right]\nonumber\\  
\chi_{1/2}^{\rm MW}(x)&=&
\frac{1}{2}\left[\prod_{n=0}^\infty(1+x^{n+1/2})-
                          \prod_{n=0}^\infty(1-x^{n+1/2})\right],  
\end{eqnarray}
which are respectively for even and odd numbers of Majorana-Weyl (MW) fermions.
The subscripts are the conformal weights of the corresponding primary fields, 
or the energies of the ground state in each sector. In the twisted sector 
(periodic boundary conditions), there is only a single nonvanishing character:
\begin{equation}
\chi_{1/16}^{\rm MW}(x)=x^{1/16}\prod_{n=1}^\infty(1+x^n).
\end{equation} 
(The zero mode is omitted here, as it will be accounted for separately, as
already explained.) The constant $1/16$ in $E$ for the twisted ground
state here is the analogue of those in the different charge sectors as derived
in \S IVA, where it was mentioned that we cannot at present derive this one
directly from our zero-energy wavefunctions. $1/16$ is the conformal weight of 
the corresponding operator $\sigma$, the spin field, which twists the boundary
condition on the Majorana fermion, like the twist field $\cal S$ discussed 
in \S III. These three expressions are well-known as the Virasoro characters 
of the critical two-dimensional Ising model, as well as in other contexts. The
characters of the chiral algebra relevant to the edge states of the Pfaffian
are
\begin{eqnarray}
\chi_{r/q,{\rm even},{\rm untwisted}}^{\rm Pf}(x)
    &=& {\textstyle\frac{1}{2}}\chi_{0}^{\rm MW}(x)
       \left(\chi_{r/q}^+(x)+\chi_{r/q}^-(x)\right) 
       +{\textstyle\frac{1}{2}}\chi_{1/2}^{\rm MW}(x)
          \left(\chi_{r/q}^+(x)-\chi_{r/q}^-(x)\right)\nonumber\\
\chi_{r/q,{\rm odd},{\rm untwisted}}^{\rm Pf}(x)  
    &=& {\textstyle\frac{1}{2}}\chi_{1/2}^{\rm MW}(x)
           \left(\chi_{r/q}^+(x)+\chi_{r/q}^-(x)\right) 
       +{\textstyle\frac{1}{2}}\chi_{0}^{\rm MW}(x)
           \left(\chi_{r/q}^+(x)-\chi_{r/q}^-(x)\right) \nonumber\\
\chi_{(r+1/2)/q,{\rm twisted}}^{\rm Pf}(x) 
    &=& \chi_{1/16}^{\rm MW}(x)\chi_{(r+1/2)/q}^+(x).
\end{eqnarray}
It can easily be seen that these expressions are sums over states with the
necessary constraints on the combinations of fermion and charge states
included, apart from those that enter on combining right and left movers and
zero modes. The partition function is, finally,
\begin{equation}
{\cal Z}^{\rm Pf}(x,\overline{x})=\sum_{r=0}^{q-1}
   \left(\left|\chi_{r/q,{\rm even},{\rm untwisted}}^{\rm Pf}(x)\right|^2
  +  \left|\chi_{r/q,{\rm odd},{\rm untwisted}}^{\rm Pf}(x)\right|^2
  +  \left|\chi_{(r+1/2)/q,{\rm twisted}}^{\rm Pf}(x) \right|^2\,\right).
\end{equation}

\subsection{HR state}

As for the Pfaffian state, the edge states of the HR state on a cylinder can be
deduced almost immediately from the results for the disk. In the HR case, the
untwisted sector is found now to contain zero modes, while the twisted sector
does not. The zero mode, like the nonzero, right and left moving modes, can be
occupied by a spin up or a spin down fermion, or both. {}From these states we 
can
deduce the projection rule. It has the same form as for the Pfaffian, in all
sectors. The projection rule
requires even fermion number when no charge has been added, and that a fermion
is created or destroyed whenever a unit of charge is added to a single edge, so
that the chiral algebra includes an operator $\partial\Psi_\sigma 
e^{i\sqrt{q}\varphi}$, very similarly to the Pfaffian. We can therefore write 
down the characters without further ado. The characters for the fermions are:
\begin{eqnarray}
\chi_{0}^{\Psi}(x)&=&\frac{1}{2}\left[\prod_{n=1}^\infty(1+x^n)^2 
                                   +\prod_{n=1}^\infty(1-x^n)^2\right]
                                                             \nonumber\\
\chi_{1}^{\Psi}(x)&=&\frac{1}{2}\left[\prod_{n=1}^\infty(1+x^n)^2 
                                   -\prod_{n=1}^\infty(1-x^n)^2\right]
                                                               \nonumber\\
\chi_{-1/8}^{\Psi}(x)&=&\frac{1}{2}x^{-1/8}\left[
                               \prod_{n=0}^\infty(1+x^{n+1/2})^2 
                                   +\prod_{n=0}^\infty(1-x^{n+1/2})^2\right]
                                                                \nonumber\\
\chi_{3/8}^{\Psi}(x)&=&\frac{1}{2}x^{-1/8}\left[
                                   \prod_{n=0}^\infty(1+x^{n+1/2})^2 
                                   -\prod_{n=0}^\infty(1-x^{n+1/2})^2\right].
\end{eqnarray}
Note that the first two are the even and odd fermion number states in the
untwisted sector, omitting the zero modes, while the last two are the same 
for the twisted sector, and we have used the negative conformal weight of the 
twist fields in writing the latter. In this case we have maintained the 
distinction between even and odd fermion numbers in the sector that contains 
the zero modes, so as to exhibit its fate explicitly. (If desired, the 
following approach can also be taken for the Pfaffian state, and the 
expressions already given for the partition function can be derived in this 
manner, verifying the argument, given in words in the preceding text, that 
there are only $3q$ sectors.) The characters for the chiral algebra of the 
HR state are:
\begin{eqnarray}
\chi_{r/q,{\rm ev},{\rm untw}}^{\rm HR}(x)
      &=& {\textstyle\frac{1}{2}}\chi_{0}^{\Psi}(x)
       \left(\chi_{r/q}^+(x)+\chi_{r/q}^-(x)\right) 
       +{\textstyle\frac{1}{2}}\chi_{1}^{\Psi}(x)
          \left(\chi_{r/q}^+(x)-\chi_{r/q}^-(x)\right)\nonumber\\
\chi_{r/q,{\rm od},{\rm untw}}^{\rm HR}(x)
      &=& {\textstyle\frac{1}{2}}\chi_{1}^{\Psi}(x)
       \left(\chi_{r/q}^+(x)+\chi_{r/q}^-(x)\right) 
       +{\textstyle\frac{1}{2}}\chi_{0}^{\Psi}(x)
          \left(\chi_{r/q}^+(x)-\chi_{r/q}^-(x)\right)\nonumber\\
\chi_{(r+1/2)/q,{\rm ev},{\rm tw}}^{\rm HR}(x)
      &=& {\textstyle\frac{1}{2}}\chi_{-1/8}^{\Psi}(x)
       \left(\chi_{(r+1/2)/q}^+(x)+\chi_{(r+1/2)/q}^-(x)\right)\nonumber\\ 
     & &\mbox{}+{\textstyle\frac{1}{2}}\chi_{3/8}^{\Psi}(x)
          \left(\chi_{(r+1/2)/q}^+(x)-\chi_{(r+1/2)/q}^-(x)\right)\nonumber\\
\chi_{(r+1/2)/q,{\rm od},{\rm tw}}^{\rm HR}(x)
      &=& {\textstyle\frac{1}{2}}\chi_{3/8}^{\Psi}(x)
       \left(\chi_{(r+1/2)/q}^+(x)+\chi_{(r+1/2)/q}^-(x)\right)\nonumber\\ 
     & &\mbox{}+{\textstyle\frac{1}{2}}\chi_{-1/8}^{\Psi}(x)
          \left(\chi_{(r+1/2)/q}^+(x)-\chi_{(r+1/2)/q}^-(x)\right).
\end{eqnarray}
We may now form the partition function, by combining the sectors subject to the
rules already mentioned. In particular, in the untwisted sector where the 
zero mode occurs, we may combine right- and left- moving sectors of the same
parity, in which case the zero mode may be either unoccupied or doubly
occupied, or we may combine sectors of opposite parity if the zero mode is
occupied once, which may be with either spin. Thus we find for the partition
function:
\begin{eqnarray}
\lefteqn{{\cal Z}^{\rm HR}(x,\overline{x})=} \nonumber\\
    &  &\sum_{r=0}^{q-1}\left[
               2\left|\chi_{r/q,{\rm ev}, {\rm untw}}^{\rm HR}(x)
                                                              \right|^2
              +2\left|\chi_{r/q,{\rm od}, {\rm untw}}^{\rm HR}(x) 
                                                               \right|^2 
              +2\left(\chi_{r/q,{\rm ev}, {\rm untw}}^{\rm HR}(x)
                  \overline{\chi_{r/q,{\rm od}, {\rm untw}}^{\rm HR}(x)}
                                                  \right.\right.\nonumber\\
       & & \qquad\left.\mbox{}+\left.\chi_{r/q,{\rm od}, {\rm untw}}^{\rm HR}(x)
                  \overline{\chi_{r/q,{\rm ev}, {\rm untw}}^{\rm HR}(x)}
                                                                   \right)
               +\left|\chi_{(r+1/2)/q,{\rm ev},{\rm tw}}^{\rm HR}(x)
                                                                \right|^2
              +\left|\chi_{(r+1/2)/q,{\rm od}, {\rm tw}}^{\rm HR}(x)
                                                                \right|^2
                                                            \right]\nonumber\\
    &=&\sum_{r=0}^{q-1}\left(2\left|
           \chi_{r/q,{\rm ev}, {\rm untw}}^{\rm HR}(x)
          +\chi_{r/q,{\rm od}, {\rm untw}}^{\rm HR}(x) 
                                                     \right|^2
        +\left|\chi_{(r+1/2)/q,{\rm ev},{\rm tw}}^{\rm HR}(x)
                                                           \right|^2
            +\left|\chi_{(r+1/2)/q,{\rm od}, {\rm tw}}^{\rm HR}(x)
                                                                \right|^2
                                  \right)\nonumber\\
    & &\mbox{}
\end{eqnarray}
which shows that there are in fact $4q$ sectors. The untwisted characters have
combined into simpler ones, similarly to the twisted Pfaffian state characters:
\begin{equation}
\chi_{r/q,{\rm untw}}^{\rm HR}(x)=
              \prod_{n=1}^\infty(1+x^n)^2\chi_{r/q}^+(x),
\end{equation}
which, however, appear twice in $\cal Z$. The $4q$ sectors show that there are
$4q$ primary fields of the chiral algebra in the system, which (in the notation
of \S III, except that the fields $\Psi_\sigma$, $\cal S$ are now the 
nonchiral fields that act on both right and left movers) are $1$ (the
identity operator), $\Psi_\uparrow\Psi_\downarrow$,
${\cal S}e^{i(\varphi+\overline{\varphi})/2\sqrt{q}}$,
$\hat{\cal S}_{\sigma\overline{\sigma}}e^{i(\varphi+\overline{\varphi})
/2\sqrt{q}}$ ($\hat{\cal S}_{\sigma\overline{\sigma}}$ is the twist field
excited in both left and right sectors, and $\sigma$,
$\overline{\sigma}=\uparrow$, $\downarrow$), 
and these operators times additional factors 
$e^{i(\varphi+\overline{\varphi})/\sqrt{q}}$ which shift charge $1/q$ from one 
edge to the other. All of these fields are spin singlets, except 
$\hat{\cal S}_{\sigma\overline{\sigma}}$ which transforms as spin ${1\over
2}\otimes{1\over2}=0\oplus 1$.
The field $\Psi_\uparrow\Psi_\downarrow$ is not strictly a primary field since
it has weight zero, but this is not an important distinction here.
All other fields are descendants of these, that
is they can be obtained by acting with operators in the chiral algebras; as a
particular example, the state created from the untwisted ground state with
unoccupied zero modes by the zero mode of $\Psi_\uparrow$ times a unit charge
at the right moving edge is obtained from $\Psi_\uparrow\Psi_\downarrow$ by
acting with $\partial\Psi_\uparrow e^{i\sqrt{q}\varphi}$. Acting again, with 
a similar operator, leads us to the identity, so the representations are not
irreducible, which is a peculiarity of this system.

\subsection{331 state and the hierarchy and its generalizations}

The edge states and the partition function for the 331 state are, by now,
easily obtained. There are zero modes in the twisted sector, as for the
Pfaffian, but there are two types of fermions (particles and antiparticles) as
in the HR state. Similar selection rules governing even and odd fermion numbers
apply as in the other cases. Accordingly, the partition function can be written
using the Weyl (or chiral Dirac) characters:
\begin{eqnarray}
\chi_{0}^{\rm Weyl}(x)&=&\frac{1}{2}\left[
                               \prod_{n=0}^\infty(1+x^{n+1/2})^2 
                                   +\prod_{n=0}^\infty(1-x^{n+1/2})^2\right]
                                                                \nonumber\\   
\chi_{1/2}^{\rm Weyl}(x)&=&\frac{1}{2}\left[
                                   \prod_{n=0}^\infty(1+x^{n+1/2})^2 
                                   -\prod_{n=0}^\infty(1-x^{n+1/2})^2\right]
                                                                  \nonumber\\
\chi_{1/8}^{\rm Weyl}(x)&=&\frac{1}{2}x^{1/8}\left[\prod_{n=1}^\infty(1+x^n)^2 
                                   +\prod_{n=1}^\infty(1-x^n)^2\right]
                                                             \nonumber\\   
\chi_{9/8}^{\rm Weyl}(x)&=&\frac{1}{2}x^{1/8}\left[\prod_{n=1}^\infty(1+x^n)^2 
                                   -\prod_{n=1}^\infty(1-x^n)^2\right].
\end{eqnarray}
The characters entering the partition function are:
\begin{eqnarray}
\chi_{r/q,{\rm ev},{\rm untw}}^{331}(x)
      &=& {\textstyle\frac{1}{2}}\chi_{0}^{\rm Weyl}(x)
       \left(\chi_{r/q}^+(x)+\chi_{r/q}^-(x)\right) 
       +{\textstyle\frac{1}{2}}\chi_{1/2}^{\rm Weyl}(x)
          \left(\chi_{r/q}^+(x)-\chi_{r/q}^-(x)\right)\nonumber\\
\chi_{r/q,{\rm od},{\rm untw}}^{331}(x)
      &=& {\textstyle\frac{1}{2}}\chi_{1/2}^{\rm Weyl}(x)
       \left(\chi_{r/q}^+(x)+\chi_{r/q}^-(x)\right) 
       +{\textstyle\frac{1}{2}}\chi_{0}^{\rm Weyl}(x)
          \left(\chi_{r/q}^+(x)-\chi_{r/q}^-(x)\right)\nonumber\\
\chi_{(r+1/2)/q,{\rm ev},{\rm tw}}^{331}(x)
      &=& {\textstyle\frac{1}{2}}\chi_{1/8}^{\rm Weyl}(x)
       \left(\chi_{(r+1/2)/q}^+(x)+\chi_{(r+1/2)/q}^-(x)\right)\nonumber\\ 
     & &\mbox{}+{\textstyle\frac{1}{2}}\chi_{9/8}^{\rm Weyl}(x)
          \left(\chi_{(r+1/2)/q}^+(x)-\chi_{(r+1/2)/q}^-(x)\right)\nonumber\\
\chi_{(r+1/2)/q,{\rm od},{\rm tw}}^{331}(x)
      &=& {\textstyle\frac{1}{2}}\chi_{9/8}^{\rm Weyl}(x)
       \left(\chi_{(r+1/2)/q}^+(x)+\chi_{(r+1/2)/q}^-(x)\right)\nonumber\\ 
     & &\mbox{}+{\textstyle\frac{1}{2}}\chi_{1/8}^{\rm Weyl}(x)
          \left(\chi_{(r+1/2)/q}^+(x)-\chi_{(r+1/2)/q}^-(x)\right).
\end{eqnarray}
The partition function is
\begin{eqnarray}
\lefteqn{{\cal Z}^{331}(x,\overline{x})=} \nonumber\\
    & &\sum_{r=0}^{q-1}\left(\left|\chi_{r/q,{\rm ev},{\rm untw}}^{331}(x)
                                                           \right|^2
            +\left|\chi_{r/q,{\rm od}, {\rm untw}}^{331}(x)
                                                           \right|^2  
           +2\left|\chi_{(r+1/2)/q,{\rm ev}, {\rm tw}}^{331}(x)
                     +\chi_{(r+1/2)/q,{\rm od}, {\rm tw}}^{331}(x) 
                                                          \right|^2
                                                     \right).\nonumber\\
     & & \mbox{}
\end{eqnarray}
Again the twisted terms have combined to form a simpler character,
\begin{equation}
\chi_{(r+1/2)/q,{\rm tw}}^{331}(x)=
(\chi_{1/16}^{\rm MW}(x))^2 \chi_{(r+1/2)/q}^+(x), 
\end{equation}
and there are two distinct sectors with this 
character. The equality of some characters of distinct sectors may also happen 
with the rational torus or Laughlin state characters, for which the characters 
obey $\chi_{r/q}^\pm(x)=\pm\chi_{(q-r)/q}^\pm(x)$. 

By bosonization \cite{stone}, the Dirac (or Weyl) characters can be written, 
using the Jacobi triple product formula, in terms of characters for a chiral 
boson with $q=1$ (summed over charge sectors): 
\begin{eqnarray}
\chi_{0}^{\rm Weyl}(x)\pm\chi_{1/2}^{\rm Weyl}(x) &\equiv& \chi_{0/1}^\pm(x)  \\
2\left(\chi_{1/16}^{\rm MW}(x)\right)^2  &\equiv& \chi_{(1/2)/1}^+(x) 
\end{eqnarray}
and so the 331 partition functions can be written in the form of sums for two
boson fields, which is described in detail below.
As mentioned in \S II, the bosonized description of the field theories is 
closely related to the description of the bulk wavefunctions as two-component 
generalizations of the Laughlin states. The $E$'s for the various sectors can 
be obtained in that description by an argument similar to that given in \S IVA 
for the Laughlin state. Since this includes a contribution from the fermions 
as well as from the charge degrees of freedom, this holds out some hope that a 
derivation in the ``pairing'' representation of wavefunctions, which might 
also be applicable to the Pfaffian and HR states, should exist.

Here we will give without proof the general results for the hierarchy and its
generalizations, restricting ourselves, for simplicity, to the case where the 
matrix $G$ below is positive definite, with the 331 state as a special case.
Physically, this is the case where all modes at the same edge propagate in the
same direction. The other case has been discussed recently in
\cite{kanefish,hald95}.
The bosonized field theory for the right-moving edge \cite{wen} can be 
formulated in terms of chiral boson fields $\varphi_\alpha$, which have 
correlators (in imaginary time)
\begin{equation}
\langle\varphi_\alpha(z)\varphi_\beta(0)\rangle=-\delta_{\alpha\beta}\ln z
\end{equation}
where $\alpha$, $\beta$ run from $1$ to $n$, and in the hierarchy $n$ is here
the number of levels. The $\alpha=1$ component could be taken to be the density
fluctuation field we have used up to now. The others represent the internal,
neutral degrees of freedom. In the composite fermion approach, 
$n$ is the number of Landau levels for the fermions \cite{read90}, and the
density mode is usually taken to be the sum of the $\varphi_\alpha$. The 
chiral operators that are
allowed to be used at this edge without affecting the other (which generate 
the chiral algebra) are of the form $e^{i v_\alpha \varphi_\alpha (z)}$ (we 
use the summation convention). The vectors $\bf v$, whose components in the 
basis labelled by $\alpha$ are $v_\alpha$, lie on a (Bravais) lattice 
$\Lambda$ in $n$-dimensional Euclidean space, that is they take the form of 
integral linear combinations of $n$ linearly independent vectors ${\bf e}_a$. 
Thus the chiral algebra can be generated by $e^{\pm ie_{a\alpha}
\varphi_\alpha(z)}$, $a=1$, \ldots, $n$.
The scalar products of the ${\bf e}_a$ are $G_{ab}={\bf e}_a\cdot{\bf e}_b$, 
which defines the Gram matrix of $\Lambda$; $G$ is positive definite here 
because we assumed Euclidian space. In the (generalized) hierarchy 
theory, $G$ is a matrix of integers, and so $\Lambda$ is an integral lattice. 
If ${\bf v}=v_a{\bf e}_a$ then we have ${\bf v}\cdot{\bf v}'=v_a G_{ab}v_b'$. 
Since the basis labelled by $\alpha$ (which is not an integral basis) is 
orthonormal, we find that the conformal weight of the chiral operators is 
${\bf v}^2/2$, and so is either integral or half-integral. 

The possible shifts of charge or the other U(1) quantum numbers related to the
components of $\varphi_\alpha$ from one edge to the other are described by 
similar operators that act on both edges simultaneously, as in the states
considered earlier. The right-moving part of such an operator is of the form 
$e^{iw_\alpha\varphi_\alpha(z)}$, where ${\bf w}$ is a vector in the dual (or
reciprocal) lattice $\Lambda^\ast$ of $\Lambda$; the dual lattice is defined as
the set of all vectors $\bf w$ such that 
${\bf w}\cdot{\bf v}=$ an integer for all ${\bf v}\in\Lambda$. Clearly
$\Lambda$ is a sublattice of $\Lambda^\ast$. The right-moving conformal weight 
(or ground state energy in the corresponding charge sector) of the operators 
$e^{iw_\alpha\varphi_\alpha(z)}$ is again ${\bf w}^2/2$, which is the same 
(modulo $1$) as the statistical parameter $\theta/2\pi$ of the quasiparticles 
in the bulk of the same state, which are also labelled by vectors in 
$\Lambda^\ast$ (the ``excitation lattice'' in the terminology of \cite{read90}).
The lattice $\Lambda$ (the ``condensate lattice'') labels the combinations of
quasiparticles that make up the possible order parameters. The relation of
these to changes in the charges at one edge was mentioned in \cite{read90}, and
forms the basis for the results quoted here. If we view the lattices as
additive groups, we find that $\Lambda$ has index $\det G$ in $\Lambda^\ast$,
and so the quotient group $\Lambda^\ast/\Lambda$ is a finite group of $\det G$ 
elements. The extended charge sectors are labelled by the possible shifts 
modulo fields in the chiral algebra, that is, by the elements of 
$\Lambda^\ast/\Lambda$. These can be described by a set of vectors 
${\bf w}_A$, $A=1$, \ldots, $\det G$, in $\Lambda^\ast$, one in each coset of 
$\Lambda$. The simplest case is the Laughlin states, where $n=1$, $G=(q)$,
$\Lambda=\sqrt{q}{\bf Z}$, $\Lambda^\ast={\bf Z}/\sqrt{q}$, and
$\Lambda^\ast/\Lambda={\bf Z}_q$, which is equivalent to the description
in \S IVA. Another simple case is the integral quantum Hall effect, in which
for $n$ Landau levels, $G$ is (in a convenient basis) the $n\times n$ 
identity matrix, so $\Lambda$ is the $n$-dimensional simple (hyper-) cubic 
lattice, $\det G=1$, and $\Lambda^\ast=\Lambda$. Thus, in this case, all 
edge excitations are just electrons in the various Landau levels. 

For the 331 states, the matrix $G$ in the basis natural for the ground state
wavefunction in the form (\ref{mmnground}) is
\begin{equation}
G^{331}=\left(\begin{array}{cc}
                q+1&q-1\\
                q-1&q+1
                  \end{array}\right).
\end{equation}
The resulting lattice $\Lambda$ embodies the projection rules by being distinct
from an orthogonal direct sum of one dimensional lattices. 
We note that there are $\det G=4q$ extended charge sectors, as found earlier.
Moreover, unlike the
lattices for the hierarchy states, for the generalized hierarchy states, and
the 331 state in particular, the sublattice $(\Lambda^\ast)^\perp$ of operators
that are neutral is not the same as the sublattice $\Lambda^\perp$ of 
$\Lambda$. The lowest-weight neutral operator that is in $\Lambda^\ast$ but 
not in $\Lambda$ represents the Fermi field $\psi_\sigma$ of weight 1/2. Since 
it does not appear in $\Lambda$, it cannot be applied to one edge, but must be 
combined with a similar operator for the other edge, or with another Fermi 
field or a charged field at the same edge. This is a consequence of the
projection rules that we saw using the fermionic form of wavefunctions; states
differing only by one fermion at one edge do not both exist in the Hilbert 
space. In contrast, for the hierarchy, the projection rules place no
restriction on the neutral operators that can be applied to one edge, since all 
vectors in $(\Lambda^\ast)^\perp$ are also in $\Lambda^\perp$.

We may now describe the partition function for the (generalized) hierarchy
states, in the case where $G$ is positive definite. The sectors are labelled 
by the $\det G$ cosets of $\Lambda$, and in each sector the chiral characters 
are sums over vectors in the coset, together with fluctuations in the 
$\varphi_\alpha$. That is, define
\begin{equation}
\chi_A(x)\equiv\sum_{{\bf v}\in \Lambda} x^{({\bf w}_A+{\bf v})^2/2}
\prod_{m=1}^\infty(1-x^m)^{-n}.
\end{equation}
The partition function is simply
\begin{equation}
{\cal Z}^\Lambda(x,\overline{x})=\sum_{A=1}^{\det G}|\chi_A(x)|^2.
\end{equation}
For the 331 states, this agrees with that derived by bosonization from 
${\cal Z}^{331}$.

It is interesting that, in the theory of the edge states, the relation of 
composite boson and composite fermion approaches maps exactly onto the usual 
$1+1$-dimensional bosonization (or its inverse, fermionization).

\subsection{Orbifolds, chiral superalgebras, and modular transformations}

For readers familiar with, or ready to learn about, CFT, we
mention that the theories for the cylinder described in this section are
examples of the construction known as ``orbifolding''. Definitions, results 
and examples of orbifolds can be found in Refs.\ \cite{dfms,ginsparg,dvvv}. 
In brief, the general algebraic definition of an orbifold involves starting 
with a rational CFT with a chiral algebra $\cal A$ on which 
some finite group $\cal G$ acts as a symmetry. One then takes the subalgebra 
${\cal A}_0$ that is invariant under $\cal G$ as the new chiral algebra. The 
representations of $\cal A$ will be representations of ${\cal A}_0$ also, but 
will in general be reducible; each irreducible component transforms as an 
irreducible representation of $\cal G$. In addition, there will be new 
``twisted'' 
representations of ${\cal A}_0$ that are not representations of $\cal A$. 
The same operations are applied to the left-moving chiral algebra 
${\overline{\cal A}}$ and its representations. The 
(symmetric, diagonal) orbifold CFT then has a primary 
field for each representation of ${\cal A}_0$, which at the same time are 
primary for the isomorphic left-moving algebra $\overline{\cal A}_0$. 
The rule for combining left- and right-moving representations is that
all fields must be invariant under the simultaneous action of $\cal G$ on left 
and right movers, and untwisted (twisted) fields must combine with untwisted
(twisted). 

The CFTs of all the paired states described in this section
are examples of orbifolds with ${\cal G}={\bf Z}_2$. The algebra $\cal A$ for 
the Pfaffian is generated by the fields $e^{\pm i\sqrt{q}\varphi}$, $\psi$, 
and contains the U(1) current algebra generated by $\partial\varphi$ 
and the Virasoro algebra for $\psi$ as subalgebras. The primary fields are 
$e^{ir\varphi/\sqrt{q}}$, $r=0$, $1$, \ldots, $q-1$ (we suppress the
left-moving operators for now). ${\bf Z}_2$ acts simultaneously on $\varphi$,
$\psi$ by $\varphi\rightarrow \varphi+\pi\sqrt{q}$ and $\psi\rightarrow -\psi$.
The algebra ${\cal A}_0$ is generated by $\psi e^{\pm i\sqrt{q}\varphi}$, and
the primary fields are $e^{ir\varphi/\sqrt{q}}$, $\psi e^{ir\varphi/\sqrt{q}}$,
$r=0$, $1$, \ldots, $q-1$, which result from the splitting of the
representations of $\cal A$ (the untwisted representations), together with the 
twisted representations $\sigma e^{i(r+1/2)\varphi/\sqrt{q}}$, $r=0$, $1$, 
\ldots, $q-1$, which include the spin field $\sigma$, which is the analogue 
for the Majorana fermion of the twist field discussed in \S III. The states (or 
descendant fields)
in these representations obey the ``projection rules'' found earlier, and the
full description of the combination of left and right movers, and the resulting
partition functions, can be done in agreement with the rules obtained from the 
wavefunctions in this section. Very similar descriptions work for the HR and 
331 states. (For the 331 state, the orbifold we find is that where
$\psi_\uparrow$ transforms by $e^{i\pi}$ and $\psi_\downarrow$ transforms by
$e^{-i\pi}$. These factors are both equal to $-1$, but the point is that the
factors written describe the way the phase of either field winds on taking it
round the twist field; there is also an adjoint twist field around which they
wind the reverse way. This behavior is required by the structure of the twisted
states, which have definite pseudospin as well as charge quantum numbers. It is
most easily understood in the bosonized representation $\varphi=\varphi_1$, 
$\psi_\uparrow=e^{i\varphi_2}$, in the same notation as in \S IVD. Then the 
symmetry is $\varphi_1\rightarrow \varphi_1+\pi\sqrt{q}$, 
$\varphi_2\rightarrow \varphi_2+\pi$. This orbifold leads back to the lattice
described in \S IVD. In particular, the spin (or twist) field for 
$\psi_\sigma$ is bosonized as $e^{i\varphi_2/2}$, which must appear in 
combination with some other charged fields, as for the other orbifolds and 
the generalized hierarchy theories in \S IIID.) 
In all cases, the rationale for the structure is that the electron 
(or other fundamental charged particle) is represented in the edge theory by 
a field like $\psi e^{\pm i\sqrt{q}\varphi}$, which has fixed boundary 
conditions in all sectors, and all fields must be local with respect to it, 
just as in the bulk, all wavefunctions must be single valued functions of the 
electron coordinates \cite{mr}.

Our description of the orbifolds glossed over one aspect of the systems
discussed here. The usual definition of chiral algebras assumes that all fields
in the chiral algebras (both $\cal A$ and ${\cal A}_0$) have integral conformal
weight. In our examples, $\psi$ and $\psi_\sigma$ that appear for the Pfaffian
and 331 states have half-odd-integral weight, and for $q$ odd, so does 
$e^{\pm i \sqrt{q}\varphi}$. Thus $\cal A$ is actually a chiral superalgebra in
these cases \cite{mr}, and so is ${\cal A}_0$ in some cases (and also for the
algebra of the Laughin state for $q$ odd, and the generalized hierarchy states 
whenever applicable to electrons). We emphasize that, for our purposes, a
superalgebra is one where some fields have half-odd-integral conformal weight,
rather than one where some of the relations are anticommutators instead of
commutators. In fact, to describe electrons, which are 
fermions, rather than the quantum Hall effect of charged bosons, the chiral 
algebra is always a superalgebra except in the case of the HR states, due to 
the violation of the spin statistics theorem there as discussed earlier. 

The fact that the chiral algebra is sometimes a superalgebra has consequences 
for the modular transformation properties that we may expect for the partition
functions calculated in this section. 
If $x=e^{2\pi i \tau}$, and ${\rm Im}\,\tau>0$ ($\tau$ should not be confused
with earlier uses of the same symbol), then modular 
transformations act as 
\begin{equation}
\tau\mapsto {a\tau+b \over c\tau+d}
\end{equation}
and the matrix 
\begin{equation}
\left(\begin{array}{cc}
         a&b\\
         c&d    \end{array}\right) 
\end{equation}
is a member of ${\rm SL}(2, {\bf Z})$, the group of $2\times2$ integer 
matrices of determinant $1$. (The group of modular transformations themselves 
is ${\rm SL}(2, {\bf Z})/\{\pm I\}$.) The modular group is generated by the
elements $T:\tau\rightarrow \tau+1$, represented by
\begin{equation}
T=\left(\begin{array}{cc}
         1&1\\
         0&1    \end{array}\right), 
\end{equation}
and $S:\tau\rightarrow -1/\tau$, represented by 
\begin{equation}
S=\left(\begin{array}{cc}
         0&1\\
        -1&0    \end{array}\right). 
\end{equation}

When the chiral algebra is strictly an 
algebra (i.e.\ not a superalgebra), then the partition functions will be 
modular invariant, if we modify the definition to include the factor 
$(x\overline{x})^{-c/24}$, where $c$ is the central charge of the CFT (not the
matrix element just above). 
Central charges are additive; the values of the central charge 
are $c=1$ for the Laughlin state, $1+1/2=3/2$ for the Pfaffian, $1-2=-1$ for 
the HR, $1+1=2$ for the 331 (all independent of the value of $q$), and $n$ for 
the (generalized) hierarchy states. 
Modular invariance occurs for $q$ even in the case of
the Laughlin and HR states, and $q$ odd for the Pfaffian and 331 states,
all of which except the HR state describe the fractional quantum Hall effect 
of charged bosons, not electrons. 
When some of the fields that generate the chiral algebra have half-integral 
conformal weight, they will obey an antiperiodic boundary condition in the space
direction, and the (modified) partition function cannot be invariant under the 
full modular group; the only boundary conditions that are invariant 
under the whole group are periodic around any cycle on the torus.            
In these cases, which as we have seen apply to all states considered here
that can describe electrons, with the sole exception of the HR states with $q$
even, we expect that our expressions are invariant only under the subgroup of 
the modular group that leaves the antiperiodic boundary condition on the 
electron field invariant. This subgroup is generated by the elements $S$ and 
$T^2$ and can be shown to be isomorphic to $\Gamma_0(2)/\{\pm I\}$, where 
$\Gamma_0(2)$ is the subgroup of ${\rm SL}(2,{\bf Z})$ consisting of matrices 
where the matrix element $c\equiv0$ (mod $2$).

We also mention here some isomorphisms of the chiral algebras of our systems to 
known algebras. (In this paragraph, $c$ is the central charge, and $N$ is not 
the number of particles.) For the Laughlin state (of bosons) at $\nu=1/2$, the 
fields $e^{\pm i \sqrt{2}\varphi}$, $\partial\varphi$ generate the SU(2) current
(Kac-Moody) algebra of level 1. For the Laughlin state at $\nu=1/3$, we have 
\cite{mr} the $N=2$ superconformal algebra at $k=1$, generated by 
$e^{\pm i \sqrt{3}\varphi}$, $\partial\varphi$. For the Pfaffian state (of
bosons) with $q=1$, the operators $e^{\pm i\sqrt{q}\varphi}$ are the bosonized 
representation of a Dirac field, or of a pair of Majorana fields 
$\psi_{\pm 1}$, which together with the Majorana field $\psi=\psi_0$ forms a 
triplet of Majoranas. This $c=3/2$ theory contains an SU(2) current algebra of 
level 2, or equivalently an O(3) algebra of level 1, in which the currents are 
the bilinears $\psi_a\psi_b$, $a$, $b=\pm 1$, $0$. This symmetry shows up, for 
example, in the degeneracies of the excited energy levels, as long as the 
velocities for $\varphi$ and $\psi$ are equal. The $3q=3$ sectors, even
untwisted, odd untwisted, and twisted, correspond to 
primary fields that transform respectively as spins 0, 1, 1/2, under both
the left- and right-moving SU(2). Moreover, the product
$\psi_1\psi_0\psi_{-1}$ generates $N=1$ superconformal symmetry \cite{dgh}, 
though this operator does not survive the projection to ${\cal A}_0$. Finally, 
the algebra ${\cal A}_0$ for the $\nu=1/2$ Pfaffian state is generated by 
$\psi e^{\pm i\sqrt{2}\varphi}$, which has weight $3/2$, and the algebra can 
be recognized as superconformal $N=2$ at $k=2$ \cite{dgh}. In this case, the 
unprojected algebra $\cal A$ contains SU(2) level 1 and an SU(2) triplet of 
supercurrents $\psi e^{\pm i\sqrt{2}\varphi}$, $\psi \partial\varphi$, which 
generate an $N=3$ superconformal algebra \cite{dgh}.

\section{Conclusion}

To conclude, we have found complete descriptions of the wavefunctions, the 
Hilbert spaces and the field theories of the edge states of the paired systems 
considered. The explicit wavefunctions are very appealing and make the 
enumeration of excited states in terms of elementary excitations 
straightforward. The combination of complete proofs of some results, and 
enumeration for low excited states in others, makes the correctness of those 
results not explicitly proven here almost certain. The results confirm the 
general prediction in \cite{mr} of a relation between bulk and edge properties.
For the Pfaffian and HR states, this provides indirect evidence for the 
prediction of nonabelian statistics of the quasiparticles in the bulk of these 
states. The twist fields in the edge conformal field theories proposed here 
certainly have such properties when exchanged in spacetime at the edge 
(``monodromy''). For the 331 state, as for all generalized hierarchy states, 
the monodromy, and the statistics of the bulk quasiparticles, is abelian. 

The explicit wavefunctions for the edge excitations, particularly for two edges
on a cylinder, are reminiscent of results for integrable one-dimensional
systems, especially those of the Calogero-Sutherland (CS) type \cite{cs}
for which there are explicit, simple wavefunctions for the ground state and
many excited energy eigenstates. Indeed, the similarity of the Laughlin state 
and the ground state of the CS model has often been remarked. In 
the limit $L\gg N$, the Laughlin state on the cylinder essentially becomes the 
Calogero-Sutherland ground state \cite{rezhaldedge}; the coordinate transverse 
to the edges can be viewed as the canonical momentum. The edge excitations of
the Laughlin state are in one-one correspondence with the excitations of the 
CS model, and the low-energy CFT of the latter is once again a Luttinger 
liquid. It is interesting to speculate that there might be some integrable 
one-dimensional Hamiltonians with long-range interactions, generalizing the 
CS model, for which the ground and excited states might be related in a similar
way to the wavefunctions discussed in this paper. If so, then we expect the 
low-energy field theories of the one-dimensional models to be the ${\bf Z}_2$ 
orbifolds discussed in \S IV.

Finally, we note that the approach used here can be applied to other 
states for which the ground state is the zero-energy eigenstate of a suitable
local Hamiltonian, as here. An example is another paired state, the permanent 
state \cite{mr}, which is a spin singlet, and is the densest zero-energy 
eigenstate of a certain 3-body Hamiltonian \cite{rr}. The resulting theory is 
a ${\bf Z}_2$ orbifold containing spin 1/2 bosons of conformal weight 1/2 at 
the edge. Such a system, like the HR state, violates the spin-statistics 
connection, so the edge field theory is not conformal. The corresponding 
non-positive conformal field theory, whose correlators reproduce the bulk 
wavefunctions, is the $\beta$-$\gamma$ ghost system \cite{mr}, so the relation 
of bulk and edge theories is maintained.  

\acknowledgements
N.R. thanks Andreas Ludwig and Greg Moore for discussions. Research was 
supported by NSF grant no.\ DMR-91-57484.

\appendix

\section{Zero Energy States for the Three-Body and Hollow-Core Hamiltonians}

In this Appendix we will justify directly the general form of the zero energy 
eigenstates of (\ref{pfaff3bodH}), and by extension its analogues for $q>1$, 
and show in
particular that they lead to the forms for the edge states in
(\ref{ansatzpfedge}), (\ref{ansatzpfedge+qh}). We then briefly address similar
questions for the hollow-core Hamiltonian for which the HR state is the unique
ground state, and corresponding issues on the cylinder.

The Hamiltonian (\ref{pfaff3bodH}), taken with Bose statistics for the 
particles so that the
Pfaffian state with $q=1$ is a possible ground state, implies that the
wavefunction of a zero energy state vanishes whenever any three (or more)
particles coincide. This implies that zero energy states can be written in the
form of the Vandermonde determinant $\prod_{i<j}(z_i-z_j)$, times an 
antisymmetric
function that, as a function of any two coordinates $z_i$, $z_j$, may have
a simple pole at $z_i=z_j$, times the usual gaussian factors. Such a state will
have zero energy provided the antisymmetric function involved does not have a
triple pole of the form 
$$
[(z_i-z_j)(z_j-z_k)(z_k-z_i)]^{-1}
$$
as $i$, $j$, $k$ approach one another, for any $i$, $j$, $k$. A form
like 
$$
[(z_i-z_j)(z_i-z_k)]^{-1}
$$
cannot appear either, because it is symmetric in $j$ and $k$, while a form
$$
(z_j-z_k)(z_i-z_j)^{-1}(z_i-z_k)^{-1}
$$
could, but this can be rewritten as a difference of simple poles
$$
(z_i-z_j)^{-1}-(z_i-z_k)^{-1}.
$$
So all possible functions can be written as linear combinations of the 
forms already given, where the                                        
singularities involve disjoint pairs of particles. 

Without loss of generality, the general state can be taken to be a linear
combination of states written by antisymmetrizing a function obtained by 
dividing the particles into pairs, writing an odd factor for each pair 
and symmetrizing over exchange of pairs. These
conditions are of course sufficient but not necessary for the final
antisymmetrization over all particles to be nonvanishing. That is, neglecting
the omnipresent factor $\prod(z_i-z_j){\rm exp}(-\frac{1}{4}\sum|z_i|^2)$,
we must have
\begin{equation}
\sum_{\sigma\in S_N} {\rm sgn}\,\sigma
\frac
{\sum_{\tau\in S_{N/2}}\prod_{k=1}^{N/2}f_k(z_{\sigma(2\tau(k)-1)},
z_{\sigma(2\tau(k))})}
{ (z_{\sigma(1)}-z_{\sigma(2)})\cdots(z_{\sigma(N-1)}-z_{\sigma(N)}) }
\label{genzeroen}
\end{equation}
where the $f_k$ are symmetric polynomials in two variables. For $N$ odd we can
write a similar form with $k=1$, \ldots, $(N-1)/2$, and include for the unpaired
particle an arbitrary polynomial factor $f_0(z_{\sigma(N)})$.

A convenient way to describe the symmetric functions in two variables $z_1$,
$z_2$, is the  following. We know that symmetric functions can be written as
sums of products of the sums of powers $s_n^{(2)}=z_1^n+z_2^n$. We will separate
the symmetric functions that vanish at $z_1=z_2$ by writing the disjoint sets
of functions 
\begin{equation}
A_m=\{(z_1-z_2)^{2m}s_n^{(2)}:n=0,1,2,\ldots\}
\label{defsetsfns}
\end{equation} 
for $m=0$, $1$, $2$, \ldots. We claim that the full set of symmetric functions
in two variables is spanned by linear combinations of the polynomials in the
set
\begin{equation}
\bigcup_{m=0}^\infty A_m
\end{equation}
(there is no need to take products of these functions). This can be shown by
induction from the fact that products of sums of powers span the symmetric
polynomials, together with the identities
\begin{eqnarray}
s_{n_1}^{(2)} s_{n_2}^{(2)}&=& 2s_{n_1+n_2}^{(2)} 
-(z_1^{n_1}-z_2^{n_1})(z_1^{n_2}-z_2^{n_2}) \label{sopident1}\\
z_1^n-z_2^n&=&(z_1-z_2)(z_1^{n-1}+z_1^{n-2}z_2+\ldots+z_2^{n-1})
\label{sopident2}
\end{eqnarray}
which (by induction on the order) express a product of elements of $A_0$ as a
linear combination of elements of $A_0$, $A_1$, \ldots.

Now each $f_k$ in (\ref{genzeroen}) can be chosen to be an element of
$\bigcup_{m=0}^\infty A_m$. If $f_k$ is an element $s_n^{(2)}$ of $A_0$, we
will try to pull outside the sum on permutations $\sigma$ the corresponding
sum of powers in all $N$ coordinates, $s_n$ (see eq.\ (\ref{sn})); this will 
leave behind terms with fewer $f_k\in A_0$. Repeating this procedure, 
eventually all $f_k$'s remaining inside the sum will be in 
$\bigcup_{m=1}^\infty A_m$ and we will have finished. $f_k$ that are in $A_m$ 
($m\geq 1$) contain $(z_{\sigma(2\tau(k)-1)}-z_{\sigma(2\tau(k))})^2$ which 
cancels a factor in the denominator, so these particles are unpaired in this 
term and the wavefunction will be a linear combination of the forms
(\ref{ansatzpfedge}). 

For the basic (untwisted) sector of edge states, we can consider $N$ large and 
most $f_k=1$, though this is not necessary and the results below are valid for 
all wavefunctions of the stated form. Then we observe that if $f_1=s_n^{(2)}\in
A_0$, then
\begin{equation}
\sum_{\tau\in S_{N/2}}\prod_{k=1}^{N/2}f_k
\propto s_n\sum_{\tau\in S_{N/2}}\prod_{k=2}^{N/2}f_k
- 
\sum_{\tau\in S_{N/2}}\sum_{k=2}^{N/2}
\hat{f}_k
\prod_{k'=2,k'\neq k}^{N/2}f_{k'}
\label{sopmanip} 
\end{equation}
where for $k=2$, $3$, \ldots, 
\begin{equation}
\hat{f}_k=\left\{\begin{array}{ll}
                   0 & \mbox{if $f_k=1$}\\
                   f_k s_n^{(2)} & \mbox{if $f_k\neq 1$,}
                  \end{array}
          \right.
\end{equation}
and $s_n$ can be taken outside the sum on permutations $\sigma$. The functions 
$f_k s_n^{(2)}$ can then be reduced using the identity (\ref{sopident1}), 
and all the terms in the many-particle state are now of the form of symmetric
polynomials in $N$ variables times antisymmetric functions with fewer $f_k$
that are members of $A_0$ and $\neq 1$. Eventually, all $f_k$ are either $1$ or
are $\in A_m$ ($m\geq 1$), and these states are linear combinations of the 
states in the text. Similar methods work for $N$ odd. Thus we have shown that 
all zero energy states are linear combinations of symmetric polynomials times 
the form in (\ref{ansatzpfedge}).

To obtain the twisted sector we can replace $f_k$ by $f_k s_1^{(2)}$ in the
above proof, leaving the $s_1^{(2)}$ factors intact inside the sum on $\sigma$
and $\tau$ at each step. Of course, in a finite system, our proof shows that
these states can be expressed as combinations of the others, but to study the
Hilbert spaces of edge states, we take $N\rightarrow\infty$ before the 
number of $f_k\neq 1$ becomes large, and thus we obtain two different sectors
in this limit. Similar arguments apply if it is desired to include any other
factor in every $f_k$ in the state.

The hollow-core Hamiltonian \cite{hr} requires that zero energy states have no
pairs of particles with relative angular momentum $q-1$. (Another way to say
this, which is useful in other geometries, is in terms of the order of
vanishing of the functions.) We recall that the
relative angular momentum of a pair, say $1$, $2$, is defined by expressing the
wavefunction in the form (neglecting the gaussian factor, and the spin labels
if any)
\begin{equation}
\Psi(z_1,z_2,z_3,\ldots,z_N)=\sum_{m,n=0}^\infty (z_1-z_2)^m(z_1+z_2)^n
        \Psi_{mn}(z_3,\ldots,z_N)
\end{equation}
in which each term in the sum is an eigenstate of relative angular momentum of
$1$ and $2$ of eigenvalue $m$. 
The densest zero energy state of the hollow-core Hamiltonian occurs at filling 
factor $\nu=1/q$. The largest $q$ for which the pairing
in which we are interested can occur is $q=2$. For $q>2$, zero energy states
can be obtained from those for $q=2$ by multiplying by the Vandermonde
determinant. For $q=2$, the wavefunctions are required to be totally 
antisymmetric when the spin states are included (see \S IIC).
For fixed spins of the $N$ particles, the wavefunctions can be written as
$\prod_{i<j}(z_i-z_j)^2$ times a meromorphic function, and the meromorphic
function must be antisymmetric among particles of the same spin. For zero 
energy states this function must have, as any two particles come to the same 
point, either a double pole, with zero residue, or be analytic. 
Because of antisymmetry, double poles can appear only for opposite spin 
particles. All antisymmetric functions can be obtained by antisymmetrization of
functions of indefinite symmetry, though we may as well omit functions that
would vanish on antisymmetrization. If a double pole is present for a pair 
$i$, $j$, then it cannot be present for any other pair of the form $i$, $k$ or 
$j$, $k$. This is because the Vandermonde squared contains the factors
\begin{equation}
(z_i-z_j)^2(z_i-z_k)^2(z_j-z_k)^2
   =(z_i-z_j)^2\left\{\left[{\textstyle\frac{1}{2}}(z_i+z_j)-z_k\right]^2
         -{\textstyle\frac{1}{4}}(z_i-z_j)^2\right\}^2
\end{equation}
the expansion of which contributes only even powers to the relative angular
momentum of $i$ and $j$. In view of the double pole in 
$z_i-z_j$, the whole wavefunction is a zero energy eigenstate,
provided there is not a single or double pole in 
$z_i-z_k$ or $z_j-z_k$, for any $k$.  Therefore, all pairing factors
$(z_i-z_j)^{-2}$ must contain distinct pairs. 
The unantisymmetrized function can thus be written as a product of pair factors
for as many opposite spin pairs as possible, times functions
$f_k(z_i^\uparrow,z_j^\downarrow)$ of the paired coordinates that can be taken
either symmetric or antisymmetric, and must be either nonvanishing at 
$z_i^\uparrow=z_j^\downarrow$, or vanish at least as fast as 
$(z_i^\uparrow-z_j^\downarrow)^2$, so as not to spoil the
zero energy property. The zero energy wavefunctions thus have, without loss of 
generality, a form similar to (\ref{genzeroen}). We now try to pull any one of 
the $f_k$ that does not vanish at $z_i^\uparrow=z_j^\downarrow$ outside the 
sum on permutations by the same procedure as for the Pfaffian, using 
(\ref{sopmanip}). Use of (\ref{sopident1}), (\ref{sopident2}) then shows that 
the resulting functions still obey the zero energy property. The procedure can 
then be repeated until linear combinations of the form (\ref{ansatzhredge}) are 
reached. We conclude that the wavefunctions in the form (\ref{ansatzhredge}) 
span all the zero energy states, in the untwisted sector. Similar arguments 
apply to the twisted sector, to $N_\uparrow\neq N_\downarrow$, and to 
combinations of these. 

Finally, we comment on zero energy states on the cylinder. On replacing $z_i$
by $Z_i$ (see \S IV) we see that in (\ref{genzeroen}) $f_k$ must still be
symmetric but may contain negative powers of $Z_i$. (The pairing factors
$(Z_i-Z_j)^{-1}$ can be left unchanged without loss of generality.) 
We extend the definition of the sets of symmetric functions (\ref{defsetsfns})
by allowing the exponents $n$ in the symmetric polynomials $s_n^{(2)}$ to
be negative as well as positive or zero, while $m$ is still non-negative; 
we claim that these span all symmetric holomorphic functions in two variables
on the cylinder. (\ref{sopident1}), (\ref{sopident2}) apply unchanged to all 
integral values of $n_1$, $n_2$, although (\ref{sopident2}) becomes an infinite
series. The proof then works as before, by pulling sums of (positive or 
negative) powers outside the sum on permutations. The HR case works similarly.

\section{Linear independence for small $\Delta M$ in the Pfaffian and HR cases}

In this appendix we construct and verify the linear independence of the states
in the untwisted, even $N$ sector for all $\Delta M \leq 8$ for the Pfaffian 
and $\Delta M \leq 6$ for the HR state. We use a different basis from that 
derived in Appendix A. We first return to the two-quasihole states for the 
Pfaffian. Due to the symmetry of exchanging $w_1$ and $w_2$, they may be 
expanded in the form:
\begin{equation}
\Psi(z_{1},\ldots,z_{N};w_{1},w_{2})=\sum_{m=0}^{N/2}\sum_{n=0}^{m}\Psi_{m n}
(z_{1},\ldots,z_{N})(w_{1}^{n}w_{2}^{m-n}+w_{2}^{n}w_{1}^{m-n})
\end{equation}
where all the $\Psi_{mn}$ are linearly independent. This may be interpreted as
saying that the quasiholes behave as two bosons, which may each occupy any one
of $N/2+1$ states. This does not, however, mean that the quasiholes are bosons
in general, which would contradict the assertion that they obey nonabelian
statistics \cite{rr}. To obtain the expansion, we first expand the
numerator in (\ref{pfaff2qholes}) inside the Pfaffian, i.e.\ for a fixed 
choice of pairs, described by the permutation $\sigma$ (each pairing is 
obtained from $2^{N/2}(N/2)!$ different $\sigma$'s). For each pair 
$\sigma(2k-1)$, $\sigma(2k)$ the factor
$[(z_{\sigma(2k-1)}-w_1)(z_{\sigma(2k)}-w_2)+(w_1\leftrightarrow w_2)]$ 
will contribute $z_{\sigma(2k-1)}z_{\sigma(2k)}$, 
$(z_{\sigma(2k-1)}+z_{\sigma(2k)})$ or a constant to an expansion coefficient 
$\Psi_{mn}$. This observation suggests the use of an alternative basis for the 
space of edge states spanned by the $\Psi_{mn}$, defined by
\begin{eqnarray}
\lefteqn{\Phi_{\Delta M,s}(z_{1},\ldots,z_{N})=} \nonumber\\
 & & \frac{1}{2^{N/2}(N/2)!}\sum_{\sigma\in S_N}
\frac{{\rm sgn}\, \sigma}{(z_{\sigma(1)}-z_{\sigma(2)})\cdots
(z_{\sigma(N-1)}-z_{\sigma(N)})}
\{\{(z_{\sigma(1)}z_{\sigma(2)})^{m_{1}}\cdots \nonumber\\
& &\qquad\cdots(z_{\sigma(N-1)}z_{\sigma(N)})^{m_{N/2}}
(z_{\sigma(1)}+z_{\sigma(2)})^{n_{1}}\cdots(z_{\sigma(N-1)}+z_{\sigma(N)})
^{n_{N/2}}\}\}  \nonumber\\
& &\qquad\qquad\times\prod_{i<j}(z_{i}-z_{j})^{q}\exp[-\frac{1}{4}
\sum|z_{i}|^{2}]
\end{eqnarray}
where $\Delta M=M-M_0$ is again the difference between the total angular 
momentum $M$ of the edge state and the angular momentum of the ground state 
$M_{0}$. The expression in the double curly brackets is defined as the sum over 
permutations of $N/2$ pairs:
\begin{eqnarray}
\lefteqn{\{\{(z_{\sigma(1)}z_{\sigma(2)})^{m_{1}}\cdots
                   (z_{\sigma(N-1)}z_{\sigma(N)})^{m_{N/2}}
(z_{\sigma(1)}+z_{\sigma(2)})^{n_{1}}\cdots
            (z_{\sigma(N-1)}+z_{\sigma(N)})^{n_{N/2}}\}\}=} \nonumber\\
&&{\cal N}^{-1}\!\sum_{\tau\in S_{N/2}} (z_{\sigma(1)}
            z_{\sigma(2)})^{m_{\tau(1)}}\cdots
                  (z_{\sigma(N-1)}z_{\sigma(N)})^{m_{\tau(N/2)}}
(z_{\sigma(1)}+z_{\sigma(2)})^{n_{\tau(1)}}\cdots
                  (z_{\sigma(N-1)}+z_{\sigma(N)})^{n_{\tau(N/2)}}\nonumber\\
& & 
\label{defnotn}
\end{eqnarray}
which makes this expression invariant under permuations of the
pairs, and under permutations $n_\alpha$, 
$m_\alpha$ $\mapsto$ $n_{\tau'(\alpha)}$, $m_{\tau'(\alpha)}$.  $\cal N$ is the 
number of permutations in $S_{N/2}$ that leave the sequence of pairs 
$n_\alpha$, $m_\alpha$, $\alpha=1$, \ldots, $N/2$ invariant.
In the states $\Phi_{\Delta M,s}$ the 
numbers $n_\alpha$, $m_\alpha$ are defined to be 0 or 1, such that 
$n_\alpha+m_\alpha\leq1$, and
$\sum_{\alpha=1}^{N/2}m_\alpha=\Delta M-s$, 
$\sum_{\alpha=1}^{N/2}n_\alpha=2s-\Delta M$. With these restrctions there is
clearly just one distinct polynomial 
of the form (\ref{defnotn}) for each $s$, which will be denoted 
$P_{\Delta M,s}$, and we see that $s\leq\Delta M\leq 2s$, 
$\Delta M-s\leq N/2$, $2s-\Delta M\leq N/2$, and $s\leq N/2$. 
Comparing $\Phi_{\Delta M,s}$ and $\Psi_{mn}$, we see that $\Delta M=N-m$.

{}From (B2) it is easy to calculate how many edge states of fixed 
$\Delta M$ the
expansion of two quasiholes gives, as $N\rightarrow\infty$. There are 
$1+\Delta M/2$ linearly
independent states for $\Delta M$ even and $(\Delta M + 1)/2$ for
$\Delta M$ odd. But for fixed $\Delta M$
\begin{equation}
\sum_{s\geq\Delta M/2}^{\Delta M} 
P_{\Delta M,s}=e_{\Delta M}
\label{Prel}
\end{equation}
where $e_{\Delta M}$ is an elementary symmetric polynomial, independent of the
permutation $\sigma$, which can therefore be brought outside the sum on
permutations as a multiplicative factor.
This arises because
bringing the two quasiholes to the same position, $w_1=w_2$, produces a single 
Laughlin quasihole. The remaining edge states, which span spaces of dimensions 
$\Delta M/2$ for $\Delta M$ even and $(\Delta M-1)/2$ for $\Delta M$ 
odd, require nontrivial factors inside
the sum over permutations and represent degrees of freedom that are not simply
density fluctuations at the edge. At each $\Delta M$, the number of such states 
coincides with the number of states with two fermions added to the ground state 
and the same $\Delta M$ in the Majorana field theory. 

Further expansions of the states with more than two quasiholes should generate,
besides further symmetric polynomial factors, all even-fermion-number
excitations. Wen \cite{wen3} has demonstrated numerically for 
(\ref{pfaff3bodH}) for up to 
$N=10$ particles that the number of low $\Delta M$ zero energy states 
coincides with that in the Majorana field theory. We will go a little further 
analytically, for arbitrary $N$. States with $2n$ quasiholes, $n>1$, at 
positions $\{w_{1},\ldots,w_{2n}\}$ will, in place of the factor
\begin{equation}
\prod_{k=1}^{N/2}[(z_{\sigma(2k-1)}-w_{1})(z_{\sigma(2k)}-w_{2})
+(w_{1}\leftrightarrow w_{2})]   
\label{2qholeinsert}
\end{equation}    
inside the sum on permutations $\sigma$ in (\ref{pfaff2qholes}), have a 
{\em product} of such factors, each involving a distinct pair of $w$'s. Thus 
the degree of the wavefunction will be $N_\phi=q(N-1)-1+n$.
There are $2^n n!$ distinct ways to associate the $w$'s in pairs, but only 
$2^{n-1}$ of the resulting electron wavefunctions are linearly independent (as
functions of the $z_i$'s for fixed $w$'s) for 
$n>1$ \cite{rr}.
Provided that they are degenerate in energy, which is true by inspection for
the appropriate 3-body Hamiltonian,
the fact that this number is $>1$ is the basis for nonabelian statistics. 
When the quasiholes are exchanged adiabatically, the usual Berry phase 
is replaced by a matrix acting in this space of degenerate quasihole 
states; however, this has not yet been explicitly demonstrated in 
this or any other example (see \cite{rr}). Here we are interested to see what 
edge excitations we can obtain by expanding these states. 
Evidently expanding in powers of $w_1$, \ldots, $w_{2n}$ will generate the
general polynomials (\ref{defnotn}) inside the sum on permutations, 
without restrictions on the $m_\alpha$'s and $n_\alpha$'s.
We will then have to take into account the linear relations 
among some of these states, corresponding to those discussed in \cite{rr} for 
$n=2$.

The linear relations among some of the states are obtained from the following
general identity. For any set of complex numbers $a_i$, $i=1$, $\ldots$, $P$, 
$P>2$ even,
\begin{equation}
 {\rm Pf}\,(a_{i}-a_{j})=0.
\label{pfaffident}
\end{equation}
This follows because the Pfaffian is the square root of a determinant in which
any three rows or columns obey a linear relation. All cases with $P>4$ can be
viewed as applications of the identity for $P=4$.
One consequence of the identity is that when we insert the expression
\begin{equation}
\{\{(z_{\sigma(1)}-z_{\sigma(2)})^{2}(z_{\sigma(3)}-z_{\sigma(4)})^{2} \}\}
\label{vanfact}
\end{equation}
(defined in analogy with (\ref{defnotn})),
or similar expressions, into the Pfaffian the resulting
expression vanishes. (\ref{vanfact}) can be expanded in the basis
(\ref{defnotn}) and 
this
gives a linear relation among the states obtained. Therefore for
$\Delta M=4$ we ``lose'' one state. For $\Delta M=5$ the
expressions are
\begin{equation}
\{\{(z_{\sigma(1)}-z_{\sigma(2)})^{2}(z_{\sigma(3)}-z_{\sigma(4)})^{2}
(z_{\sigma(5)}+z_{\sigma(6)})\}\}
\end{equation}
\begin{equation}
\{\{(z_{\sigma(1)}-z_{\sigma(2)})^{2}(z_{\sigma(3)}-z_{\sigma(4)})
^{2}\}\}e_1.
\end{equation}
Notice that in
the first expression the distinct pair $\sigma(5)$, $\sigma(6)$ is introduced,
which does not affect the vanishing which is due to the summation over
permutations of the other four particles. In the second we have simply
multiplied the insertion (\ref{vanfact}) by an elementary symmetric polynomial,
which is linearly independent of the other expression. From here on we will 
omit the expressions which are products 
of the ones valid for lower $\Delta M$ and symmetric polynomials.
Then for $\Delta M=6$ the linearly independent expressions producing linear
relations among the states are:
\begin{equation}
\{\{(z_{\sigma(1)}-z_{\sigma(2)})^{2}(z_{\sigma(3)}-z_{\sigma(4)})^{2}
(z_{\sigma(5)}-z_{\sigma(6)})^{2}\}\},
\end{equation}
\begin{equation}     
\{\{(z_{\sigma(1)}-z_{\sigma(2)})^{2}(z_{\sigma(3)}-z_{\sigma(4)})^{2}
(z_{\sigma(5)}+z_{\sigma(6)})^{2}\}\},
\end{equation}
\begin{equation}
\{\{(z_{\sigma(1)}-z_{\sigma(2)})^{2}(z_{\sigma(3)}-z_{\sigma(4)})^{2}
(z_{\sigma(5)}+z_{\sigma(6)})(z_{\sigma(7)}+z_{\sigma(8)})\}\}
\end{equation}
and
\begin{equation}
\{\{(z_{\sigma(1)}-z_{\sigma(2)})^{2}(z_{\sigma(1)}+z_{\sigma(2)})
(z_{\sigma(3)}-z_{\sigma(4)})^{2}(z_{\sigma(3)}+z_{\sigma(4)})\}\}.          
\end{equation}   
The last expression uses (\ref{pfaffident}) with $a_i=z_{i}^2$.
 For $\Delta M=7$:
\begin{equation} 
\{\{(z_{\sigma(1)}-z_{\sigma(2)})^{2}(z_{\sigma(3)}-z_{\sigma(4)})^{2}
(z_{\sigma(5)}-z_{\sigma(6)})^{2}(z_{\sigma(7)}+z_{\sigma(8)})\}\},
\end{equation}  
\begin{equation} 
\{\{(z_{\sigma(1)}-z_{\sigma(2)})^{2}(z_{\sigma(3)}-z_{\sigma(4)})^{2}
(z_{\sigma(5)}+z_{\sigma(6)})^{3}\}\},
\end{equation}  
\begin{equation} 
\{\{(z_{\sigma(1)}-z_{\sigma(2)})^{2}(z_{\sigma(3)}-z_{\sigma(4)})^{2}
(z_{\sigma(5)}+z_{\sigma(6)})^{2}(z_{\sigma(7)}+z_{\sigma(8)})\}\},
\end{equation}  
\begin{equation} 
\{\{(z_{\sigma(1)}-z_{\sigma(2)})^{2}(z_{\sigma(3)}-z_{\sigma(4)})^{2}
(z_{\sigma(5)}+z_{\sigma(6)})(z_{\sigma(7)}+z_{\sigma(8)})
(z_{\sigma(9)}+z_{\sigma(10)})\}\}
\end{equation}  
and
\begin{equation}       
\{\{(z_{\sigma(1)}-z_{\sigma(2)})^{2}(z_{\sigma(1)}+z_{\sigma(2)})
(z_{\sigma(3)}-z_{\sigma(4)})^{2}(z_{\sigma(3)}+z_{\sigma(4)})
(z_{\sigma(5)}+z_{\sigma(6)})\}\}.  
\end{equation}  
All these expressions are linearly independent of each other. 

Since symmetric polynomials can always be multiplied into zero energy states to
obtain another zero energy state, it is convenient to decompose all states into
a product of a symmetric polynomial and another part that is linearly
independent of symmetric polynomials. The latter represents excitations that
are not density fluctuations at the edge. The full Hilbert space of edge 
excitations thus can be written as a tensor product of a bosonic Fock space of 
density excitations, as described earlier, and another space of independent
excitations. Since the $\Delta M$s of the excitations add, the dimension of 
the full space at any $\Delta M$ can be obtained by convoluting those of the
two factor spaces. It is easy to calculate the dimensions obtained for the
latter space by building its states up from products of the $P_{\Delta M, s}$
(rendered linearly independent of $e_{\Delta M}$) and then subtracting the
number of linear relations just obtained.
For $\Delta M\leq 7$, we find that the linear relations eliminate all the
states obtained from more than two quasiholes.
Thus we find that for $\Delta M\leq 7$ the edge excitations of the
Pfaffian state exactly match those in the chiral boson times Majorana fermion 
system, in the fermion number zero or two sectors. 

In principle, it is possible to find
the number of the edge states at arbitrarily high $\Delta M$, by
deriving these expressions in a systematic way. First we list all 
polynomials of degree $\Delta M$ of the form
\begin{equation}       
\{\{(z_{\sigma(1)}-z_{\sigma(2)})^{2}
(z_{\sigma(3)}-z_{\sigma(4)})^{2} \cdots \}\}  
\label{2brokepairs}
\end{equation}
where dots denote additional squared differences or sums that multiply
the first two terms. Then we take the space of expressions that vanish when
inserted in the Pfaffian at lower $\Delta M$, multiplied with all possible 
products of symmetric polynomials that make the degree of the expression 
$\Delta M$.
We expand these in the terms of the form (\ref{2brokepairs}). Some terms in the
expansion are obviously zero when inserted in the Pfaffian; then the
rest must give zero too. Fortunately for low momenta ($\Delta M \leq 7$)
each wave function of the form (\ref{2brokepairs}) is zero and that leaves us
to prove only that the rest of the wavefunctions are non-zero and
linearly independent. This can be done by taking pairwise limits
$z_1\rightarrow z_2$, etc,
of particle coordinates in the Pfaffian alone (i.e.\ without 
Laughlin-Jastrow factor), whenever these are singular, and examining the linear
independence of the resulting functions of the remaining variables, that are
the residues of these poles.

We will just state that the number of the edge states  we found at
$\Delta M=8$ implies that the four fermion state $(\frac{1}{2},1\frac{1}{2},
2\frac{1}{2},3\frac{1}{2})$ is present in the spectrum. So the numbers
found are those given in the table in \S IIB.

We now turn to the HR state. States with any even number, $2n$, of quasiholes
are zero energy states and can serve as generating functions for edge states. As
in the Pfaffian case they suggest an overcomplete basis of the states in which
polynomials of the (\ref{defnotn}) type are inserted. 
For more than two quasiholes, linear dependences arise when more than one 
factor like $(z^\uparrow_i-z^\downarrow_j)^2$ cancels a similar factor in the
denominator; this time, the factors in the denominator are themselves squared,
and the sum over permutations of the relevant particles gives a determinant,
not a Pfaffian, so the identity that replaces (\ref{pfaffident}) is simply that
\begin{equation}
\det \left( {\begin{array} {cccc}
                           1&1&\cdots&1\\
                           1&1&      & \\
                           &\vdots&  & \\
                           1&1&      &1 
             \end{array}}\right)=0.
\end{equation}
The linear dependencies 
that we mentioned in the Pfaffian case with ($i\uparrow \sigma(i) \downarrow$)
pairs instead of ($\sigma(i)\sigma(i+1)$) are then valid here, except the 
last one
in the $\Delta M=6$ and $\Delta M=7$ cases. Namely, if we insert any of these 
in the HR state (\ref{HRground}) the sum over permutations will produce zero. 
Therefore, the numbers of nontrivial edge states, that is, the states without 
symmetric polynomials we found so far in the Haldane-Rezayi case are
\begin{center}
\begin{tabular}{c|c|c|c|c|c|c|c}
 $\Delta M$ & 1 & 2 & 3 & 4 & 5 & 6 & 7 \\ \hline
 dim       & 0 & 1 & 1 & 2 & 2 & 4 & 4
\end{tabular} \quad .
\end{center}
All these states are singlets because they consist of singlet pairs.      

The complete spectrum of edge states of the HR system should contain also 
nonzero spin states. At present we have no quasihole-like generating functions 
for these, but we can still obtain the edge states by writing down suitable 
functions directly. Each such state
is formed when some of spin-singlet pairs in the ground state are excited 
into triplet states of zero energy. The broadest class of edge states made
in this way, with $S_{z}=0$, has this polynomial in the numerator of the term
with a fixed arrangement of coordinates in pairs:
\begin{equation}
\{\{(z_{1}^{\uparrow}-z_{\sigma(1)}^{\downarrow})^{n_{1}}\cdots
(z_{N/2}^{\uparrow}-z_{\sigma(N/2)}^{\downarrow})^{n_{N/2}}
(z_{1}^{\uparrow}+z_{\sigma(1)}^{\downarrow})^{m_{1}}\cdots
(z_{N/2}^{\uparrow}+z_{\sigma(N/2)}^{\downarrow})^{m_{N/2}}\}\}
\end{equation}
where no $n_{i}=1$, and at least one of $n_{i}'s$ is
an odd number. 

For $\Delta M=3$ we have only $\{\{(z_{i}^{\uparrow}
-z_{\sigma(i)}^{\downarrow})^{3}\}\}$ which corresponds to $S=1, S_{z}=0$,
{\it  i.e.}, belongs to the permutation group representation 
$(2^{N/2-1}1^{2})$. 
Similarly the two remaining states of the triplet can be constructed. For 
example the $S_{z}=-1$ state is:
\begin{eqnarray}
\lefteqn{\Psi_{S_{z}=-1}(z_{1}^{\uparrow},\ldots,z_{N/2+1}^{\downarrow})=}
                                          \nonumber\\
 & &\sum_{\sigma\in S_{N/2+1}}{\rm sgn}\,\sigma\, 
         (z_{\sigma(1)}^{\downarrow}-z_{\sigma(2)}^{\downarrow})
            \frac{1}{(z_{1}^{\uparrow}-z_{\sigma(3)}^{\downarrow})^{2}\cdots
(z_{N/2-1}^{\uparrow}-z_{\sigma(N/2+1)}^{\downarrow})^{2}} \nonumber\\
 & & \qquad\times\prod_{i<j}(z_{i}-z_{j})^{q}\exp[-\frac{1}{4}\sum|z_{i}|^2].
\end{eqnarray}           
Note that these functions are simply related to the general form of 
(\ref{ansatzhredge}).

Then we proceed counting only linearly independent states which
do not contain symmetric polynomial factors. For low angular momenta we get 
these numbers
\begin{center}
\begin{tabular}{c|c|c|c|c|c|c}
 $\Delta M$ & 1 & 2 & 3 & 4 & 5 & 6  \\ \hline
 dim        & 0 & 0 & 1 & 1 & 2 & 2 
\end{tabular}
\end{center}
where each state in the table is $S_z=0$ element of a triplet, $S=1$. 
The total number of low-lying fermion edge states in the untwisted sector of 
the HR state is then as given in the table in \S IIC.

Finally we note that the sets of linearly independent functions obtained here
for each $\Delta M$ can be rearranged into the general form derived in Appendix
A.

\end{document}